\documentclass[twocolumn,prl,amssymb,amsmath,preprintnumbers,aps,nofootinbib]{revtex4-1}
\usepackage{lineno}
\usepackage{bm,color,xcolor}
\usepackage{slashed} 
\usepackage{graphicx}
\usepackage{soul} 
\usepackage{multirow}
\usepackage{comment}
\usepackage{epstopdf} 
\usepackage{mathtools}
\usepackage{appendix}
\usepackage[colorlinks=true
,urlcolor=blue
,anchorcolor=blue
,citecolor=blue 
,filecolor=blue
,linkcolor=blue
,menucolor=blue
,linktocpage=true
,pdfproducer=medialab
,pdfa=true
]{hyperref}

\newcommand{\beq}{\begin{equation}}
\newcommand{\eeq}{\end{equation}}
\newcommand{\bea}{\begin{eqnarray}}
\newcommand{\eea}{\end{eqnarray}}
\newcommand{\ba}{\begin{array}}
\newcommand{\ea}{\end{array}}

\def\m1{M_1}
\def\m2{M_2}
\def\m3{M_3}

\def\ch10{\tilde \chi^0_1}

\def\gev{\,{\rm GeV}}

\def\to{\rightarrow}

\newcommand{\lsim}{\mathrel{\mathop{\kern 0pt \rlap
  {\raise.2ex\hbox{$<$}}}
  \lower.9ex\hbox{\kern-.190em $\sim$}}}
\newcommand{\gsim}{\mathrel{\mathop{\kern 0pt \rlap
  {\raise.2ex\hbox{$>$}}}
  \lower.9ex\hbox{\kern-.190em $\sim$}}}
\begin{document}
\title{\boldmath \bf \Large 
The High Quality QCD Axion and the LHC
}

\author{\bf Anson Hook, Soubhik Kumar, Zhen Liu, and Raman Sundrum} 

\affiliation{Maryland Center for Fundamental Physics, Department of Physics,
University of Maryland, College Park, MD 20742, USA}
\preprint{UMD-PP-019-07}
\date{\normalsize  \today}

\begin{abstract}
The QCD axion provides an elegant solution to the Strong CP Problem. While the minimal realization is vulnerable to the so-called ``Axion Quality Problem'', we will consider a more robust realization in the presence of a mirror sector related to the Standard Model by a (softly broken) $\mathbb{Z}_2$ symmetry.
We point out that the resulting ``heavy'' axion, while satisfying all theoretical and observational constraints, has a large and uncharted parameter space which allows it to be probed at the LHC as a Long-Lived Particle (LLP).
The small defining axionic coupling to gluons results in a challenging hadronic decay signal which we argue can be distinguished against the background in such a long-lived regime, and yet, the same coupling allows for sufficient production at hadron colliders thanks to the large gluon parton luminosity. Our study opens up a new window towards accelerator observable axions, and more generally, singly produced LLPs.
\end{abstract}

\preprint{
}

\maketitle

{\it\bf Introduction.}---The Strong CP problem is the puzzle of why the strong interactions are CP symmetric even though the Standard Model (SM) as a whole is not. Technically, the question centers on the vanishingly small value of the one CP-violating coupling of QCD, $\theta$. An elegant solution is provided by elevating $\theta
$ to a fully dynamical pseudoscalar ``axion'' field. If the axion gets its potential entirely through QCD effects, then remarkably its ground state automatically corresponds to $\theta=0$~\cite{Peccei:1977hh,Peccei:1977ur,Weinberg:1977ma,Wilczek:1977pj}. 

Despite this bottom-up simplicity, the QCD axion mechanism has a top-down flaw: the axion quality problem \cite{Kamionkowski:1992mf,Barr:1992qq,GHIGNA1992278,Holman:1992us}. 
This arises because there can be other UV contributions to the axion potential that can push the minimum away from $\theta=0$. The QCD-induced potential is so shallow that even higher-dimensional interactions suppressed by UV scales, all the way up to the Planck mass, are sufficient to spoil the axion mechanism. One approach to this problem is to very strongly suppress these UV contributions with special UV structure for the axion, such as compositeness \cite{Kim:1984pt,Randall:1992ut}, extra dimensions \cite{Choi:2003wr} or string theory \cite{Svrcek:2006yi}. While such structures do ameliorate the quality problem, it is not clear to what extent they fully solve it.
However, in this paper, we present an alternative resolution.  We consider a mirror sector in the UV related to the SM by a $\mathbb{Z}_2$ symmetry \cite{Rubakov:1997vp,Berezhiani:2000gh,Hook:2014cda,Fukuda:2015ana,Dimopoulos:2016lvn}, coupled to the same axion, such that its contribution to the axion potential is much larger than QCD's but aligned with it in having its minimum at $\theta=0$. This results in a vastly higher quality axion mechanism in that it is much more robust against other uncorrelated UV effects. See Refs. \cite{Dimopoulos:1979pp,Holdom:1982ex,Dine:1986bg,Flynn:1987rs,Agrawal:2017ksf,Agrawal:2017evu,Gherghetta:2016fhp,Gaillard:2018xgk} for earlier work which also discuss other ways of having a heavy QCD axion. 

Enormous efforts are currently being devoted to the search for a standard QCD axion in the form of an extremely light and extremely weakly coupled field in small-scale experiments, astroparticle experiments, and astrophysical observations. For reviews, see Refs. \cite{Graham:2015ouw,Tanabashi:2018oca}. By contrast, we find our general solution to the quality problem places us in a very different and interesting region in axion mass-coupling parameter space, in which the axion can be probed as a quantum particle at cutting-edge collider experiments. 

The phenomenology of the high quality QCD axion should be distinguished from that of other types of light pseudoscalar particles, often dubbed axion-like-particles (ALPs) (for recent discussions on ALP effective theories see e.g. Refs.~\cite{Brivio:2017ije,Bauer:2017ris}). 
We expect that true axions will have their coupling to QCD not much weaker than other couplings to the SM. Many ALP searches/models, however, consider a strong connection and couplings to the electroweak sector, and are, therefore, not directly relevant for a true QCD axion (e.g. see Refs.~\cite{Jaeckel:2015jla,Knapen:2016moh,Izaguirre:2016dfi,Marciano:2016yhf,Brivio:2017ije,Bauer:2017ris,Dobrich:2018jyi,Craig:2018kne,Gavela:2019wzg,Merlo:2019anv,Bauer:2019gfk,Altmannshofer:2019yji}). Some ALP searches do conform to QCD axion expectation and yet couple strongly enough to the SM so that axion production and decay can be detected at colliders by traditional means, e.g., see Refs.~\cite{Jaeckel:2012yz,Mariotti:2017vtv,CidVidal:2018blh,Beacham:2019nyx,Aloni:2018vki,Alonso-Alvarez:2018irt,Ebadi:2019gij,Gavela:2019cmq}.
But we will show that a high quality axion has a significant parameter space where its coupling to QCD and the SM are so weak that its production and decays might be expected to be buried under SM background. And yet, remarkably, for \textit{sufficiently small} coupling the axion becomes a long-lived particle (LLP), spatially separable from the dominant backgrounds, and is produced in sufficient numbers because the small coupling is offset by the immense gluonic content of the colliding proton beams.
We will show that this high quality axion search presents a theoretically motivated and experimentally novel but challenging target for singly produced LLPs at the LHC main detectors. For other related on-going and proposed LLP ALP searches see Refs. \cite{Dobrich:2015jyk,Chou:2016lxi,Dolan:2017osp,Gligorov:2017nwh,Ariga:2018uku,Berlin:2018pwi,Feng:2018noy,Curtin:2018mvb,Aloni:2019ruo,Dobrich:2019dxc,Harland-Lang:2019zur}.


We begin by reviewing the axion mechanism and its quality problem, and then present our two-sector solution, which puts the axion within collider reach.
We discuss the details of the LHC phenomenology, focusing on axion production, its long-lived regime, and the suppression of the dominant background from fake-tracks. 
Finally, we discuss the results and provide our outlook.

\newpage
{\it\bf Axion mechanism and the quality problem.}---The QCD axion field, $a(x)$, is coupled to QCD by promoting $\bar{\theta}\rightarrow \bar{\theta}+a(x)/f_a$:
\beq\label{aggdual}
\mathcal{L}\supset \frac{\alpha_3}{8 \pi} \left(\bar{\theta}+\frac{a(x)}{f_a}\right) G_{\mu\nu}^a\tilde G^{a,\mu\nu}.
\eeq
In the absence of the axion, $\bar{\theta}$ represents the CP-odd gauge invariant QCD coupling, constrained by bounds on the neutron electric dipole moment to be $|\bar{\theta}|<10^{-10}$ \cite{Baker:2006ts}.
In the above, $\alpha_3=g_s^2/4\pi$ is given in terms of the QCD gauge coupling $g_s$; $\tilde{G}$ denotes the dual of the gluon field strength, $\tilde{G}^{a,\mu\nu } = \frac{1}{2}\epsilon^{\mu\nu\rho\sigma}G^a_{\rho\sigma}$ with $\epsilon^{0123}=1$; and $f_a$ denotes the axion ``decay constant''.
$\bar{\theta}$ is the \textit{effective} $\theta$-parameter obtained after diagonalizing the Yukawa matrices via chiral rotations, and is given by $\bar \theta\equiv \theta +\arg \det (Y_u Y_d)$, where
$Y_{u(d)}$ is the up (down) type Yukawa matrix with complex entries and $\theta$ is a \textit{bare} Lagrangian parameter.

The non-perturbative QCD axion potential resulting from Eq.~\eqref{aggdual}, can be calculated using chiral perturbation theory \cite{Weinberg:1977ma,DiVecchia:1980yfw},
\beq
\mathcal{V}\approx -m_\pi^2 f_\pi^2\sqrt {1-\frac {4m_u m_d} {(m_u+m_d)^2}\sin^2\left(\frac {a(x)} {2f_a}+\frac {\bar\theta} 2\right)},
\label{eq:QCDV}
\eeq
where $m_\pi$ and $f_\pi$ are respectively the mass and the decay constant of the pion, $m_{u(d)}$ is the up (down) quark mass. Given the potential in Eq.~\eqref{eq:QCDV} (and its refinements), the axion acquires a vacuum expectation value (VEV), $\langle a\rangle =-f_a \bar \theta$. Plugging this into Eq.~\eqref{aggdual} we see at low energies that the CP violation in QCD is eliminated---solving Strong CP Problem.
The mass of the axion, $m_a$, can then be obtained from Eq.~\eqref{eq:QCDV} as
\begin{equation}\label{axionmass}
m_a^2 \approx \frac{m_um_d}{(m_u+m_d)^2}\frac{m_\pi^2f_\pi^2}{f_a^2}.
\end{equation}

Clearly, if there is any other contribution to the axion potential from beyond QCD, the resulting axion ground state need no longer screen the QCD CP violation. One can realize the axion as a Nambu-Goldstone boson of a $U(1)$ Peccei-Quinn (PQ) symmetry, $e^{ia/f_a}\rightarrow e^{i\phi}e^{ia/f_a}$, which would forbid any axion potential. But such a symmetry cannot be exact because it is broken by QCD (chiral anomaly) effects, and further in the far UV 
quantum gravity is expected to explicitly break all global symmetries~\cite{PhysRevD.52.912,PhysRevD.83.084019}. At best, PQ must be an accidental symmetry of the leading couplings. For example, the UV violation of PQ symmetry may take the form of a higher-dimensional composite operator, $\mathcal{O}$, with scaling dimension $\Delta>4$ and PQ charge $q$,
\beq
\mathcal{L}\supset \frac{\mathcal{O}}{M_p^{\Delta-4}}\underset{\text{low energies}}{\sim}\frac{f^\Delta_a}{M_p^{\Delta-4}}e^{iqa/f_a},
\label{eq:planckop}
\eeq
where $M_p\sim 10^{19}$ GeV is the Planck scale. Naively, one would expect that Planck suppressed PQ violation would have negligible effects on the axion mechanism. However, given the experimental constraint $f_a>10^9$ GeV \cite{Tanabashi:2018oca} and the very delicate QCD potential Eq.~\eqref{eq:QCDV}, the axion mechanism is spoilt (for $q\sim$ 1) out to very high scaling dimension unless $\Delta\geq9$. This extreme fragility of the axion mechanism is the so-called ``quality problem'' \cite{Kamionkowski:1992mf,Barr:1992qq,GHIGNA1992278,Holman:1992us}.

While it is possible, but demanding, that the UV structure does strongly suppress PQ violation at this level, in this paper, we study a different approach: we strengthen the IR axion mechanism itself. We begin by noting that the quality problem is so severe because the QCD-induced potential realizing the axion mechanism is set by the relatively small hadronic scales $\sim 100$ MeV. This implies even smaller axion masses which could only have escaped discovery so far if the coupling $1/f_a$ to the SM is extremely small. 
We will introduce a mirror sector to the SM which reinforces the axion mechanism and mass with much larger scales, and consequently, larger couplings to the SM are experimentally allowed.

{\it\bf Model construction.}---
Here we will develop the mirror sector structure for the axion in a manner that resolves the quality problem with the minimum of other theoretical bias, leaving the broadest axion parameter space to be explored or constrained by experiments. 
We will consider a $\mathbb{Z}_2$ symmetry which exchanges between the fields of the SM and matching fields of the mirror sector. 
This replication includes the SM gauge structure, so that the entire mirror sector carries no SM charges and vice versa. All the marginal (dimensionless) couplings of the two sectors are identical, including $\bar{\theta}$. However $\mathbb{Z}_2$ is softly broken 
in the one relevant operator, by having two distinct tachyonic Higgs mass terms,
\begin{equation}
\mathcal{L}_{\text{SM}}=\mathcal{L}_{\text{kinetic}}+\mathcal{L}_{\text{marginal}}(\Psi,H,A)+\mu^2H^\dagger H
\end{equation}
\begin{equation}
\mathcal{L}_{\text{mirror}}=\mathcal{L}_{\text{kinetic}}+\mathcal{L}_{\text{marginal}}(\Psi^\prime,H^\prime,A^\prime)+\mu^{\prime 2}H^{\prime \dagger} H^\prime.
\end{equation}
We will consider $M_p^2\gg\mu^{\prime 2}\gg \mu^2$.
One can view $\mu^{\prime 2}-\mu^{2}$ as arising from the VEV of a $\mathbb{Z}_2$ odd scalar coupled to the Higgs fields in the far UV. It is then plausible that the $\mathbb{Z}_2$ breaking in the marginal couplings is as small as $\sim \mu^{\prime 2}/M_p^2$.

Let us address the theoretical plausibility of this kind of replication of SM structure from a UV perspective. For example, within a broadly string theoretic framework, it is possible that the SM gauge structure is realized at the intersection of some ``branes'' localized in an extra-dimensional manifold, see e.g. Ref.~\cite{Uranga:2007zza}. SM couplings and fields could represent the ground state and low-energy fluctuations/degeneracies of the brane arrangement in the extra dimensions. This is analogous to a molecule in ordinary space, with its ground state and (near-)degeneracies. Of course, instead of one, there can readily be two such molecules in space, made of the same atoms, and they will have identical ground states and degeneracies. It is possible that there is an environmental bias, say provided by one molecule being in a background magnetic field, in which case the two molecules are not identical. But if this bias is very weak, it is only the most fragile properties of the molecules that will differ significantly. Similarly, the SM brane arrangement may be replicated in the extra dimensions by a mirror sector with nearly identical couplings and fields. But the most significant difference is expected in the most fragile property of the SM. This is given by the mass-squared of the Higgs doublet, the very root of the Hierarchy Problem. It is plausible that the Anthropic Principle requires that one of these sectors has a small electroweak scale, but not both. This would yield the structure we have proposed above. 

We now introduce a $\mathbb{Z}_2$-invariant QCD axion coupling,
\bea\label{acoupling}
\frac{\alpha_3}{8 \pi}\left(\frac{a}{f_a}+\bar{\theta}\right) \left(G \tilde G +  G' \tilde G'\right).
\eea
We assume a $\mathbb{Z}_2$-symmetric UV completion of this non-renormalizable axionic coupling at scales $\sim f_a$, paralleling any choice of UV completion in minimal axion models without a mirror sector.
The equality of the strong couplings indicated above is true in the far UV, by $\mathbb{Z}_2$, but can run differently below $\sim\mu^\prime$.
For $\mu^\prime > 100$ TeV, $\Lambda_{\text{QCD}}\ll \Lambda_{\text{QCD}}^\prime < m_{q^\prime}$ for all mirror quarks $q^\prime$. We can estimate the strong coupling scale $\Lambda_{\text{QCD}}^\prime$ given $\Lambda_{\text{QCD}}$ using the 1-loop renormalization group. The differential running at 1-loop depends on the quark and mirror quark masses in terms of $\mu$ and $\mu^\prime$, but is insensitive to new model-dependent thresholds involving colored degrees of freedom needed to UV-complete the non-renormalizable axionic couplings, which do not get mass through electroweak symmetry breaking.

In the above regime the non-perturbative QCD$^\prime$ (pure glue) contribution to the axion potential near its minimum is given by lattice computation \cite{DelDebbio:2004ns}, and continuum ($\overline{\text{MS}}$) matching \cite{Aoki:2019cca}:
\begin{equation}\label{eq:vprime}
\mathcal{V}^\prime=0.3\times (\text{few}\times\Lambda_{\text{QCD}}^{\prime})^4 \left(\frac{a}{f_a}+\bar{\theta}\right)^2 +\mathcal{O}\left(\left(\frac{a}{f_a}+\bar{\theta}\right)^4\right).
\end{equation}
where ``few''$\sim \alpha_{3}^{-0.4}$ is in detail the model-dependent conversion between the 1-loop and 2-loop estimates of the strong coupling scales.

Considering Eq.~\eqref{acoupling}, this immediately shows that the single QCD axion $a$ solves the Strong CP problems of both the sectors at the same time by having the VEV $\langle a\rangle = -f_a\bar{\theta}$.

\begin{figure}[t]
  \centering
  \includegraphics[scale=0.30]{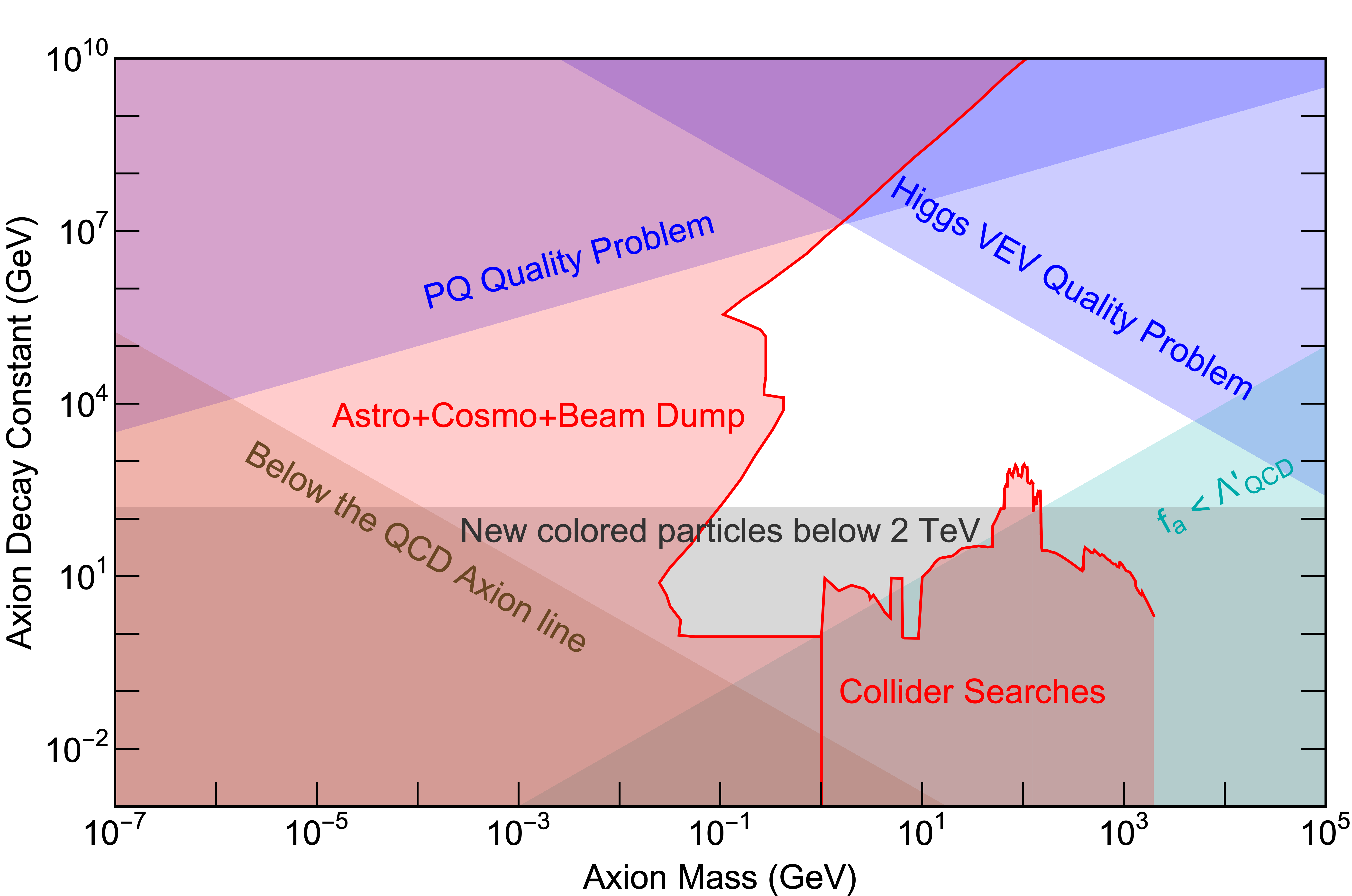}
  \caption{The preferred model parameter regions for our high quality axion model. We require that no new colored particles exist with a mass below 2 TeV; higher dimensional operators involving the axion or the Higgs do not reintroduce the strong CP problem, as well as several astrophysical, cosmological and collider constraints. We assume an axionic coupling to different gauge groups weighted by their respective fine structure constants (see Appendix \ref{app:LALP}).}
  \label{fig:model}
\end{figure}

Although the two values of $\bar{\theta}$ are identical for the two sectors in the UV, the breaking of the $\mathbb{Z}_2$ symmetry can make the two $\bar \theta$'s different below $\mu^\prime$. However, RG running of $\bar{\theta}$ occurs at seven loops, and contributions from threshold corrections arise at four loops \cite{Ellis:1978hq}. Thus both of these effects, arising from renormalizable operators in SM and SM$^\prime$, are too small to be significant even given the tight EDM constraints.

However, higher dimensional operators suppressed by the Planck scale can make the two angles different. For example, the interactions
\beq\label{higgsplanck}
\frac{\alpha_3}{8\pi}\left(\frac{H^\dagger H }{M_p^2} G \tilde G + \frac{H'^\dagger H'}{M_p^2} G' \tilde G'\right)
\eeq
can give $\bar{\theta}\neq \bar{\theta}^\prime$  upon the breaking of the $\mathbb{Z}_2$ symmetry.  Requiring that $|\bar{\theta}-\bar{\theta}^\prime|/\bar{\theta}<10^{-10}$, gives the constraint that $\mu' < 10^{14}$ GeV, and thus there is a maximal amount by which $a$ can be made heavier. Furthermore, as discussed above dimensionless couplings such as $\bar{\theta}$ can directly get a small $\mathbb{Z}_2$-breaking correction $\sim \mu^{\prime 2}/M_p^2$, which is again sufficiently suppressed for $\mu' < 10^{14}$ GeV.

From Eq.~\eqref{eq:vprime}, we see that the resulting axion mass is much larger than in the SM alone (for a given $f_a$) so that it can be heavier than $\Lambda_{\text{QCD}}$. This significantly weakens the existing experimental constraints and allows stronger couplings, $1/f_a$, to the SM. The raising of the axion mass and lowering of $f_a$ clearly reduces the severity of the quality problem. This opens up a strongly motivated and new experimentally testable regime for the QCD axion, which we identify now.

{\it\bf Constraints on parameter space.}---In Fig.~\ref{fig:model} we show the preferred parameter space for our model. We begin with the quality problem. 
We will choose as a benchmark a composite axion model for which PQ symmetry holds at the renormalizable level, but can be violated at $\Delta=6$. Given Eq.~\eqref{eq:planckop}, this reintroduces the Strong CP Problem in the region labeled as ``PQ Quality Problem'' in Fig.~\ref{fig:model}. 
We cannot populate the area labeled as ``Below the QCD Axion line'', defined by Eq.~\eqref{axionmass}, as our mechanism can only make the axion heavier, and not lighter. 
In the area labeled ``Higgs VEV Quality Problem'', $\langle H^\prime\rangle \sim \mu^\prime > 10^{14}$ GeV and Planck suppressed operators spoil the axion mechanism, as explained around Eq.~\eqref{higgsplanck}. 
Our EFT is only valid if $f_a>\Lambda_{\text{QCD}}^\prime$ which excludes the region shaded in cyan in Fig. \ref{fig:model}. 
We are being agnostic about the origin of the axion coupling to QCD.  
Typically, the coupling is generated by integrating out colored fundamental fermions who get a mass $y f_a e^{i a/f_a}$~\footnote{If the fermions were in a larger representation of $SU(3)$, then their mass could be larger by Casimir factors.}.  Requiring that the Yukawa coupling, $y$, of these fermions is smaller than $4 \pi$, we have new colored particles below $4 \pi f_a$.  Requiring that these colored fermions satisfy LHC constraints~\cite{Aaboud:2017nmi,Sirunyan:2018xlo,Sirunyan:2018rlj,Aad:2019hjw} and are heavier than $\sim$ 2 TeV
disfavors the region shown in Fig.~\ref{fig:model}. 
The region labeled ``Astro+Cosmo+Beam Dump'' is ruled out due to a variety of supernova, stellar cooling, beam dump, and cosmology constraints. The current collider coverage is shown in red regions with label  ``Collider Searches". These constraints, along with original references, can be found in Appendix~\ref{app:coverage} and Ref.~\cite{Bauer:2017ris}. Constraints shown without sharp outlines involve some of degree of theoretical uncertainty as discussed in the text.

We see then that the most favored region is given by $m_a\sim$ GeV-TeV and $f_a\sim $ TeV-PeV, ripe for collider exploration!

{\it\bf Phenomenology.}---We present a search strategy and discuss in detail its feasibility for massive axions in the GeV to tens of GeV range, with decay constants ranging between 100 TeV to PeV, thereby covering a sizable portion of the open regime seen in Fig.~\ref{fig:model}. We first note from Eq. \eqref{eq:vprime} that, for the axion to have $m_a\gtrsim$ GeV and $f_a\gtrsim$ 100 TeV, $\Lambda_{\text{QCD}}^\prime\gtrsim 100$ GeV $\gg \Lambda_{\text{QCD}}$. For this to happen in our (softly broken) $\mathbb{Z}_2$ symmetric set-up, we need $\mu^\prime > 10^{11}$ GeV.

Three crucial ingredients lead to the plausibility of realizing a massive axion search at the LHC. First, the signal is displaced, giving us a powerful discriminator against hadronic backgrounds. Second, the signal still has a sizable production rate in the displaced parameter regime. This is non-trivial for the GeV scale axion, or in general, for singly produced long-lived particles, given that the same small coupling controls the production rate and upper limit of the proper lifetime. The lever-arm that offsets the small production coupling is provided by the immense gluon parton luminosity at the LHC and other proton accelerator experiments. The third ingredient is the possibility of a low-level \textit{Displaced Track Trigger} \cite{Gershtein:2017tsv,CMS-PAS-FTR-18-018} that can be configured towards our low-mass signal, where traditional high-$p_T$ triggers would fail.

Let us see the quantitative connection between the production rate and lifetime.
The production rate of the axion is,
\beq
\sigma (pp\to a+X; H_T > 100~\gev)\simeq \left(\frac {100~{\rm TeV}} {f_a}\right)^2\times 0.7~{\rm fb}.
\label{eq:rate}
\eeq
Here we have imposed an $H_T > 100$ GeV cut and hence the axion mass ($\lesssim$ 20 GeV) does not significantly affect the cross section. 
Similarly, if one instead requires a leading jet minimal $p_T$ cut of 30~GeV, the cross section increases by around a factor of 3.
The lifetime of the axion can be approximated by,
\beq\label{eq:lifetime}
c\tau\simeq 8 \left(\frac {f_a} {\rm PeV}\right)^2 \left(\frac {2~\rm GeV} {m_a}\right)^3 {\rm cm}.
\eeq
The detailed discussion of the axion production and decays can be found in the Appendices \ref{app:production} and \ref{app:decay} respectively. We note here that axion dominantly decays into hadronic final states, and hence, with the production mode we consider the signal will be a displaced jet recoiling against prompt jet(s).

The challenges and opportunities for this signal reside in both the trigger level and post-trigger analysis. A good signal trigger efficiency is critical given the rate in Eq.~\eqref{eq:rate}, which can be achieved using the
{\it Displaced Track Trigger} discussed in Ref.~\cite{Gershtein:2017tsv,CMS-PAS-FTR-18-018}. We follow this construction with conservative modifications to accommodate our signal. In detail, we require
\begin{itemize}
  \item At least three tracks (within an L1 jet) with $p_T>2~\gev$;
  \item Amongst the above tracks, at least three of them have the transverse impact parameter $d_0 > 1$ mm;
  \item The pseudo-rapidity of the tracks to be $|\eta|<2.4$;
  \item The signal decay location in the transverse plane, $d_T < 35$ cm to have enough hits in the tracker outer layers;
  \item The $H_T$ of the event to be greater than 100~GeV.\footnote{We also consider an alternative more optimistic trigger requirement of a minimal leading jet $p_T$ of 30~GeV when showing and discussing the results. This choice increases the background cross section by about a factor of 8 while the signal is enhanced by approximately a factor of 3. A MET+displaced track trigger can also be useful.}
\end{itemize}
These requirements are sufficient to pass the Level-1 (L1) trigger with affordable rates below $\lesssim$ 10 kHz, where this is dominated by backgrounds from fake tracks \cite{Gershtein:2017tsv,CMS-PAS-FTR-18-018}. The requirement of three or more tracks can be passed quite easily for axion masses $\gtrsim$ few GeV. For example, given a fixed proper lifetime of 3 cm, whereas for $m_a=2$ GeV, 17\% of the axion decays produce three or more tracks, for $m_a\gtrsim 4$ GeV, $>$90\% of the axion decays do the same. More details about the kinematic distributions of tracks coming from axion decay can be found in the Appendix in Fig.~\ref{fig:nooftracks}.

The SM background consists mainly of metastable SM particles, such as B mesons, kaons, pions, and tau leptons. The kaon and pion have rather long lifetimes but only produce two displaced tracks. Our requirement of three displaced tracks will veto these backgrounds very effectively. For backgrounds from B mesons and tau leptons, their proper lifetimes are around 500 microns and 87 microns, respectively, producing tracks with low impact parameters. Our trigger requirement $d_0>1$ mm and requirement of a reconstructed vertex displaced by more than 5 mm vetoes them effectively. For detailed simulation and the background counting, see the Appendix.

The fake-track background is one of the most crucial for displaced vertex searches at the LHC at every stage of experimental analysis. This background comes from misconnections of the tracker hits or instrumental noises. The key feature of these fake tracks is that they allow for a much larger reconstructed impact parameter than the SM background. Empirically, one can model them as tracks with flat distributions in the finite range in the following dimensions: track impact parameter ($|d_0|<15~$cm), track curvature ($1/R\propto q/p_T < 1/(1.8~\rm m)$), track eta $|\eta|<2.4$, track time $t_0$ ($|t_0|<6$~ns) and track z-coordinate ($|z_0|<15$~cm) \footnote{These L1 fake-track parameters are mainly based upon \cite{Gershtein:2019dhy}. We note that these assumptions are rough and can vary based upon detector performance and collider environments, however, they can serve as a good benchmark to start the discussion of such LLP search strategies.} \footnote{If we assume that the above quantities are Gaussian-distributed instead, the fake-track background will decrease even further since the dominant background contribution of the fake tracks will come from the ``tail'' region.}.

Nevertheless, the fake track backgrounds that pass the above L1 triggers produce huge background, amounting to about
\beq
10~{\rm kHz} \times 10^8~{\rm sec} = 10^{12}
\eeq
triples of fake tracks at the HL-LHC.
At higher level triggers and in the analysis, one needs to suppress the background much further, to below a Hz while maintaining a high signal efficiency. Beyond all existing studies, we not only consider the issue of triggering on our signal, but we also demonstrate that it is possible to suppress these backgrounds using a 2D-4D displaced vertexing selection at high level.

The 2D-4D vertexing is defined as the following. We first solve for a 2D vertex by finding the best-fit point that minimizes the distances between the vertex and the tracks. Then we construct the 4D vertex of the system by extrapolating the 2D point in the transverse plane to the $z$ direction and time direction by propagating the tracks.
\footnote{We conservatively assume that the track momentum direction follows the direction that minimizes the flight time between the 2D impact point to the fitted vertex.} \footnote{Note that this is not directly a 4D vertex fitting that minimizes the $\chi^2$ of the 4D information of the tracks for a common vertex. This definition is to reduce the computation needed for the vertex finding. While we will show below that our 2D-4D vertexing will suppress the background to a minimal level, a full 4D fit will certainly make the result very robust providing higher background suppression.} 


\begin{figure}[htbp]
  \centering
  \includegraphics[scale=0.5]{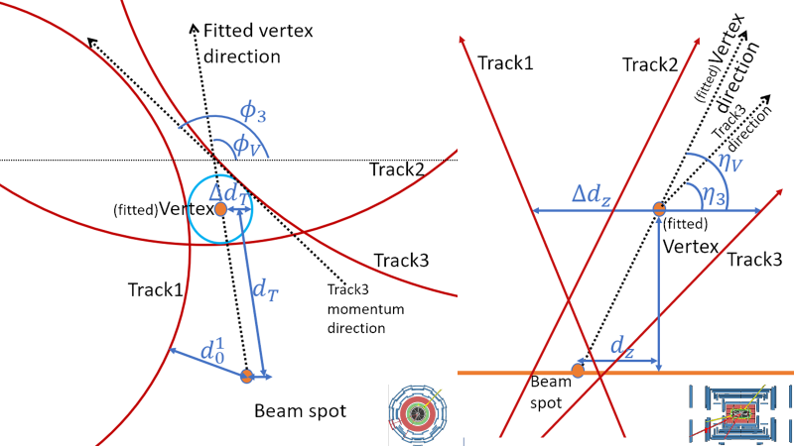}
  \caption{Schematic display of the kinematical variables used in our 2D-4D vertexing selections. The red lines represent (fake) tracks and the black dotted lines represent the reconstructed momentum and vertices.}
  \label{fig:vertexing}
\end{figure}
 

Our 2D-4D displaced vertexing selection is defined as follows:
\begin{enumerate}
\item The 2D common vertex has a minimal distance to the interaction point of 0.5~cm and maximal distance of 35~cm, $0.5~{\rm cm}<d_T<35$~cm;
\item The 2D tracks fit a common vertex with standard deviation $\Delta d_{T}<1$~cm;
\item The 2D common vertex is sufficiently displaced away from the interaction point, $d_T/\Delta d_{T}>5$;
\item The corresponding 4D vertex has a standard deviation in $z$ direction $\Delta d_z<5$~cm;
\item The corresponding 4D vertex has a $z$-direction location $d_z<20$~cm;
\item The corresponding 4D vertex has a standard deviation in time $\Delta d_t<500$~ps;
\item The corresponding 4D vertex has a time $d_t<1000$~ps;
\item The tracks are within 0.4 in pseudorapidity of the reconstructed displaced jet direction $|\eta_i -\eta_V| < 0.4$ for all the three tracks;
\item The tracks are within 0.4 in azimuthal angle of the reconstructed displaced jet direction $|\phi_i -\phi_V| < 0.4$ for all the three tracks;
\end{enumerate}

The definitions of these quantities are shown in the schematic drawing in Fig.~\ref{fig:vertexing}. Following the empirical model of fake track distribution discussed above, we find that the combination of the transverse plane vertex fitting (Cut1+Cut2+Cut3) provides a suppression factor of $8.2\times 10^{-2}$. The combination of $z$-direction consistency (Cut4+Cut5) and $t$-direction consistency (Cut6+Cut7) provides $4.9\times 10^{-2}$ and $3.0\times 10^{-3}$ background suppression, respectively. Furthermore, the requirement for the displaced tracks pointing back to the primary vertex (Cut8+Cut9) provides $4.9\times 10^{-4}$ suppression. After taking into account the correlations between the selection cuts\footnote{See supplemental material, especially the discussion on the correlations around Table~\ref{tab:correlation}, for the detailed treatment and estimation of the selection cuts correlations.}, the resulting overall suppression factor of the fake-track background from this 2D-4D vertex fitting procedure is $2.9*10^{-9}$. This means the background is reduced to
\beq
10^{12}\times 2.9*10^{-9}=2900.
\eeq
A crucial consideration on top of the above background estimation is that so far, it is using outer layers of the tracking information only. For the signal, there would be consistent energy deposition in the Electromagnetic Calorimeter (ECal) and Hadronic Calorimeter (HCal), as well as inner tracker information, which will improve the spatial resolution of the displaced tracks and constitute
a powerful consistency check. If one further requires the matching of the information between different subdetectors for all the tracks and as well the neutral hadrons, each of three tracks should be able to at least provide one order of magnitude fake-track background suppression\footnote{We thank Yuri Gershtein for private communications to confirm this point.
}, reducing our background estimate to,
\beq
2900\times (10^{-1})^3\simeq 3.
\eeq
Hence, it is plausible that the fake track background can be suppressed to negligible levels. 

Admittedly, there is large uncertainty in the fake-track background estimation. For instance, depending on the detector performance at the HL-LHC, one can have 10-30 fake tracks per collision and, the ranges in the fake track distribution model may vary. The study in Ref.~\cite{CMS-PAS-FTR-18-018} showed that by requiring two ``high quality'' fake tracks, one can have $\sim 10^{-1}$ background rate suppression and the $H_T$ cut provides $\sim 10^{-2}$ suppression, so the overall rate at L1 is around 10~kHz. 
Our evaluation here conservatively assumes that the same level of suppression can be achieved by requiring three ``high quality'' fake tracks.
As shown in the cross-section discussion in the Appendix in Ref.~\ref{app:production}, such a $H_T$ cut reduces the inclusive cross section by more than two orders of magnitude for the axion mass regions of interest. To show what one can achieve with a slightly less conservative trigger consideration, in the next section, we also consider a trigger with a leading jet $p_T$ of 30~GeV plus three ``high quality'' displaced tracks. We assume the same level of L1 rate and background can be maintained using advanced trigger developments such as matching information between different subdetectors.
Additionally, there are backgrounds from two fake tracks plus one real track, or one fake-track plus two real tracks. These backgrounds can be vetoed effectively by requiring the real tracks to have sizable impact parameters, which is already part of our preselection and trigger considerations.

Although it was not included in the 2D-4D vertexing selection above, a requirement on vertex mass will certainly help to further reduce SM background for axion masses bigger than $\sim$5-10 GeV. For smaller axion masses, a simple mass cut will cut away signals, so a more careful treatment might be needed. There is another crucial machine-related background, which is the secondary vertices created by SM particles interacting with detector materials. Naively, these backgrounds will be very similar to our signal given it has a true displaced vertex with a small mass, whose momenta also points back to the primary vertex with high probability. Special treatments, such as removing displaced vertices that originated from material-dense areas, have been applied in various displaced vertex searches at the LHC~\cite{Aaboud:2017iio,Sirunyan:2018icq,Sirunyan:2018vlw,Alimena:2019zri}. It is not yet clear how efficient this removal procedure would be for low mass searches such as ours, but it is important to remove the backgrounds from material interactions. This challenge deserves further study.  

Lastly, there may be another approach to start looking for our signals, which is through a modification of the hadronic tau tagger to exploit the large displacement. Note the similarities between our signal and the three-prong tau decays: both of them are hadronic, involve GeV scale masses, and are displaced. One could start conducting our search by imposing a further displacement requirement for a three-prong tau-like object, after removing the invariant mass cut. It would be an interesting and practical approach to start exploring our signals.

{\it\bf Results.}---
\begin{figure}[tb]
  \centering
  \includegraphics[scale=0.515]{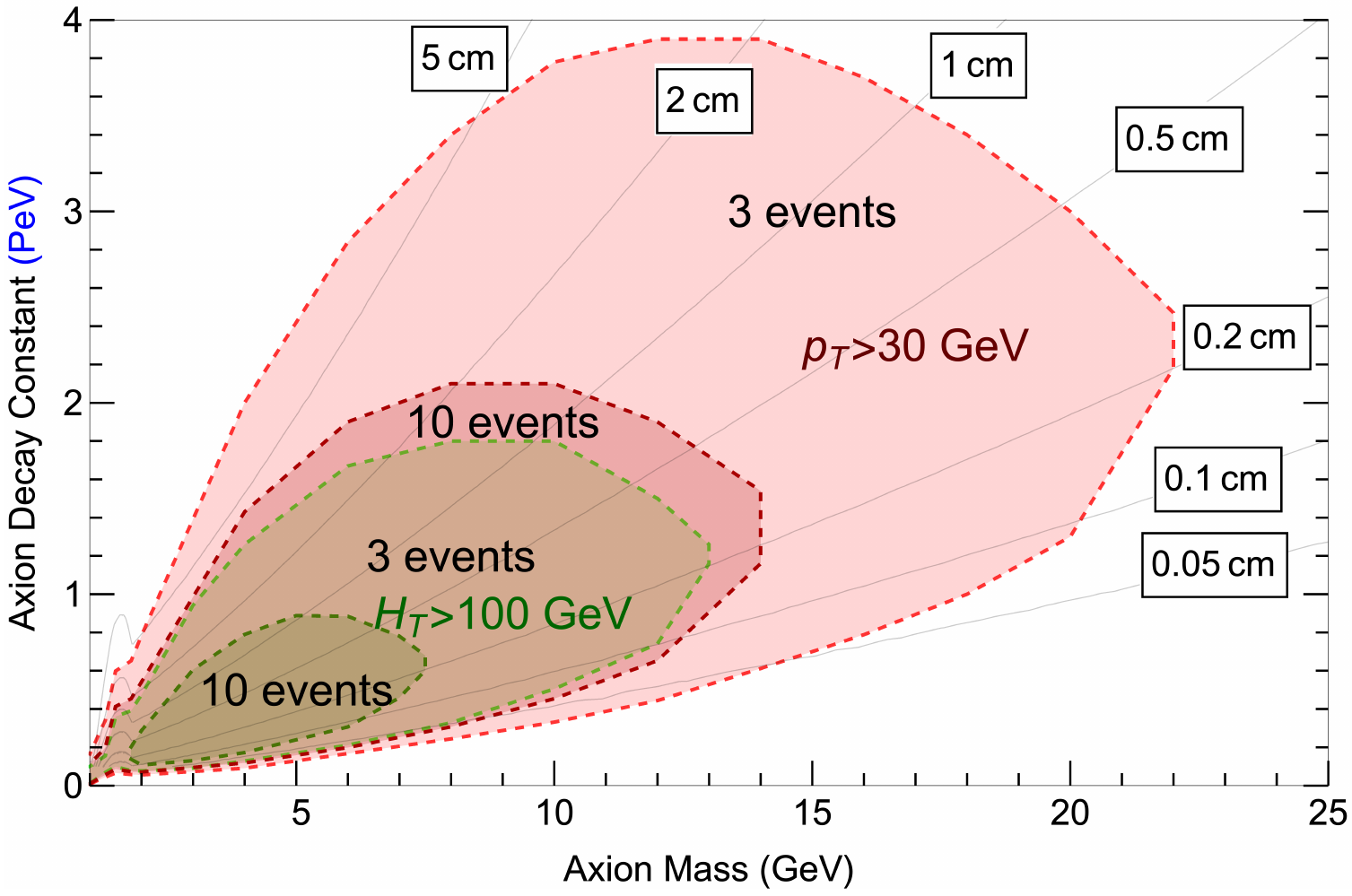}
  \includegraphics[scale=0.38]{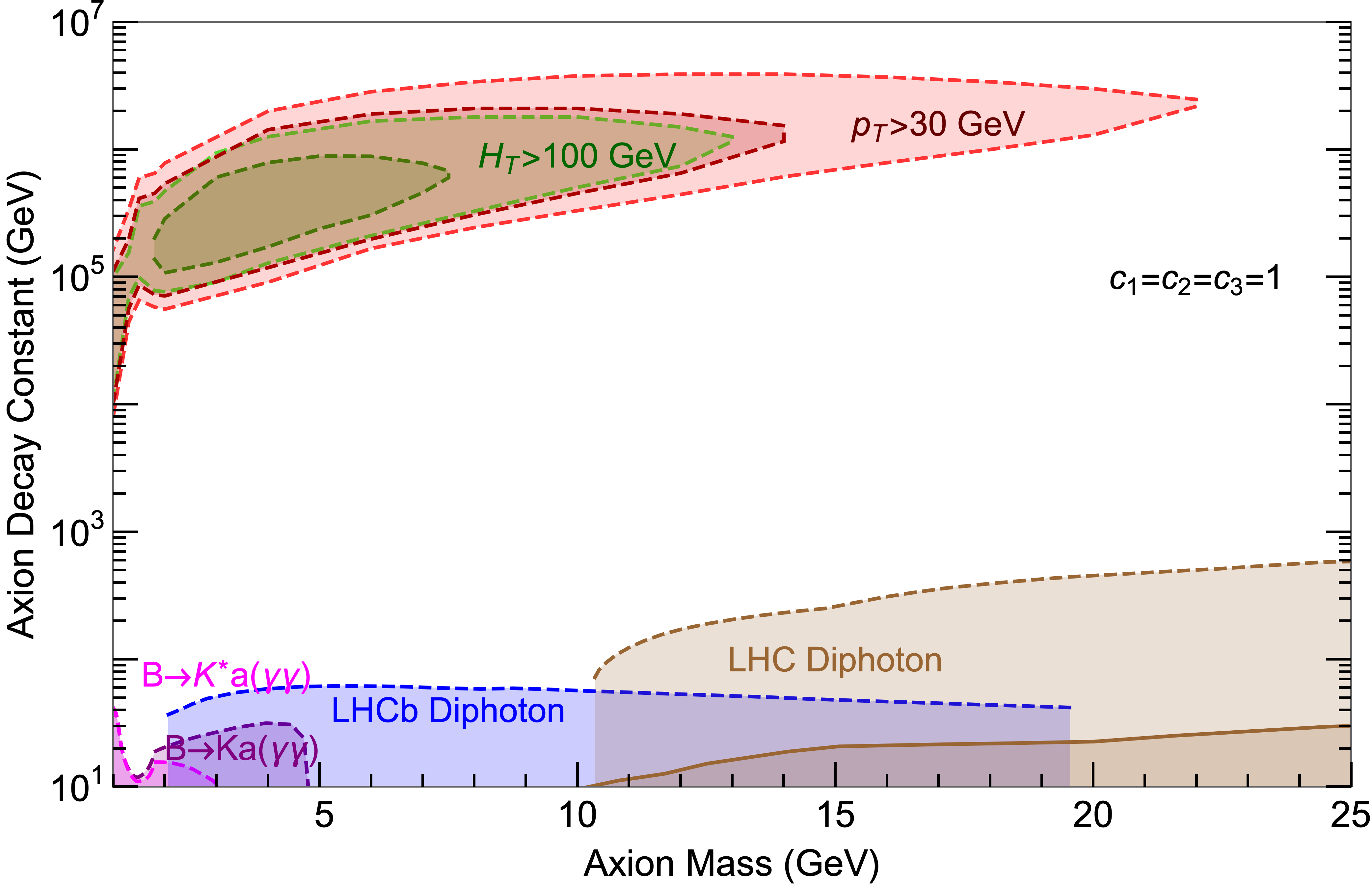}
  \caption{The signal coverage in the $m_a$ and $f_a$ plane where the red and green contours correspond to requiring a leading jet with $p_T>30$ GeV and $H_T>$ 100 GeV respectively. For both the red and green contours, we show 10 and 3 signal events regions (corresponding to 95\% C.L. exclusion limits for 25 and 0 background events respectively) that can be covered following the present work. 
  {\bf Upper panel:} The reference lines for various ALP proper lifetimes are shown.
  {\bf Lower panel:} The signal coverage in comparison with bounds (solid) and projections (dashed) coming from other (proposed) searches (for more details see Appendix \ref{app:coverage}). 
  }
  \label{fig:results}
\end{figure}
After these considerations, we show the estimated sensitivities for our signals at the HL-LHC with $3000\text{ fb}^{-1}$ integrated luminosity. The typical signal efficiencies with our selection cuts are between $10^{-3}$ to $10^{-1}$ as can be seen from Fig. \ref{fig:efflifetime}. Given the large uncertainties in background estimation, we show in Fig.~\ref{fig:results} the reach of the proposed search
with two different assumptions on the number of signal events, namely with 3 and 10 signal events that correspond to 0 and 25 background events respectively. While these numbers for the background events are just our crude estimates as described above, the LHC experiments would have refined background estimates based on full simulations and data driven methods. 

From Fig.~\ref{fig:results} we can see that with a minimal $H_T$ cut of 100~GeV (shown in green shaded region), our estimates imply a search for a displaced hadronic vertex would probe axion masses $\text{GeV}\lesssim m_a\lesssim 12\text{ GeV}$, with decay constants $100\text{ TeV}\lesssim f_a\lesssim 2 \text{ PeV}$ when 3 signal events are required. The coverage shrinks when 10 signal events are required, and then such a search would probe $\text{GeV}\lesssim m_a\lesssim 7\text{ GeV}$ and $100\text{ TeV}\lesssim f_a\lesssim 1 \text{ PeV}$. With a minimal leading jet $p_T$ cut of 30~GeV (shown in the red shaded region), the 10 events search would cover a similar region as the $H_T>100$~GeV, 3 events region, while the 3 events region extends to around $m_a\sim 20$~GeV and $f_a\lesssim 4$~PeV showing the unique potential of the main detectors in probing massive axions.

There are several unique features of the coverage plot in Fig. \ref{fig:results}, which are different from the analogous results involving more common LLP searches where LLPs are pair-produced. Understanding this will help in designing and optimizing future searches for GeV scale axions and, more generally, singly produced LLPs. First, the coverage has a strong dependence on the number of signal events needed, which is clearly shown by comparing the coverage between the shaded regions with different colors. The reason behind the strong dependence comes from our trigger requirement $H_T >100$~GeV or leading jet $p_T>30$~GeV. Given that the axion mass we are probing is much smaller than these scales, the production cross section remains approximately constant, as discussed around Eq.~\eqref{eq:rate} and shown clearly in the Appendix in Fig.~\eqref{fig:xs}. Second, the lower edge of the coverage is determined by the minimal displacement requirement, below which the efficiencies becomes too low for a GeV axion to be sufficiently displaced. Third, unlike most LLPs that are produced by stronger interactions than those involved in their decays, the upper edge for our search is limited by the production rate of the LLPs. These features echo the challenge of the search---one needs to have good background control to reach the GeV axion. We have to reduce the background as much as possible while maintaining a good signal selection efficiency. Of course, alternative signal trigger and selection strategies that do not penalize the low mass, small displacement, axions could potentially enhance coverage. This deserves further study.

The allowed region for the high quality QCD axion in Fig. \ref{fig:model} also includes a regime of relatively low decay constants $\sim 100$ GeV, which allows for copious production and prompt decays. A significant number of ongoing ALP searches at various accelerator facilities are covering this regime. 
The coverage of these searches, after translating into our axion EFT normalization,
\begin{equation}
  \frac{a}{8\pi f_a}\left(c_3 \alpha_3  G\Tilde{G}+c_2\alpha_2  W\Tilde{W}+c_1\alpha_1 B\Tilde{B}\right),
\end{equation}
are typically for $f_a\lesssim 100$ GeV. The best projected sensitivity for $m_a\sim 10$ GeV masses is the HL-LHC prompt diphoton resonance search~\cite{CidVidal:2018blh} which reaches $f_a\lesssim $ TeV. We also show our result in logarithmic scale in the lower panel of Fig.~\ref{fig:results} to compare with these existing and other proposed prompt searches, whose details can be found in the Appendix~\ref{app:coverage}. In the regime $f_a\lesssim 100$ GeV, the non-renormalizable EFT is expected to break down at energies $\lesssim 4\pi f_a$, and new SM-charged UV degrees of freedom should appear roughly within range of the LHC, providing alternate search channels, with current constraints schematically indicated in Fig.~\ref{fig:results}. But for high decay constants that we have studied, it may well be that
the axion production/detection is the \textit{only} channel available at collider energies.


{\it\bf Discussion and Outlook.}---
The quest for axions is pressing. We have put forward a general theoretical structure involving a mirror sector in which a true QCD axion solves the Strong CP problem while being very robust against the axion quality problem. We find that $\sim$ GeV axions with $\sim$ PeV decay constants lie at the heart of the motivated parameter space (see Fig.~\ref{fig:model}). 

Cosmologically, the mirror sector is constrained because any massive stable mirror states would constitute a component of dark matter and must have consistent properties and abundance, while the massless mirror photon is subject to $\Delta N_{\text{eff}}$ constraints on relativistic species. We will take a simple attitude here, and assume that the reheating temperature $T_{\text{RH}}$ after cosmic inflation is below $f_a \sim$ PeV so that all massive stable mirror states are not reheated.  Mirror glueballs are the lightest massive mirror states and may be reheated, but decay promptly into SM gluons via off-shell axion exchange. 
The mirror photon equilibrates with the $\sim 100$ degrees of freedom of the SM through axion exchange which implies $\Delta N_{\text{eff}}\sim 0.05$, which is below the current constraint of $\Delta N_{\text{eff}}\sim 0.2$ \cite{Aghanim:2018eyx}, but testable in future experiments.
For Majorana neutrinos in both sectors, originating from (even roughly) $\mathbb{Z}_2$-symmetric dimension-5 couplings to the (mirror) Higgs,  $m_{\nu^\prime}\sim \frac{\mu^{\prime 2}}{\mu^2} m_{\nu}>10^8$ GeV for $\mu^\prime > 10^{11}$ GeV and therefore the mirror neutrinos are not reheated and  pose no cosmological problem.


For $\sim$ GeV axions with $\sim$ PeV decay constants, the axion can be produced and detected at the LHC as a long-lived particle. 
This is a challenging long-lived particle search at the LHC. The production rate is highly suppressed by the same small coupling that leads to the displaced decay, implying that for reasonable rates, most decays will take place inside the LHC main detectors.
Furthermore, there is only one low-mass displaced vertex in the event, while most other searches are for pair production of massive long-lived particles. We explored the dominant background of fake tracks and proposed a three-displaced-track strategy with 2D-4D displaced vertex reconstruction, demonstrating that the background can be feasibly suppressed. We believe that this makes the case for experimental exploration.

While we have mostly focused on QCD axion motivations and modeling in this paper, there is a second general and important motivation for our proposed LLP search following from the principle of electroweak Naturalness. This broad principle gives strong reasons to expect substantial new TeV-scale physics beyond the SM in order to stabilize the Higgs sector under quantum effects. However,  we have not yet seen any evidence for such new physics, either in direct collider searches or in indirect tests of flavor physics and fundamental symmetries. This suggests that at best Naturalness is ``frustrated” by other UV or anthropic considerations or mechanisms which postpone the new physics to scales beyond the reach of our most sensitive existing probes. It is entirely plausible that the new physics scale of frustrated Naturalness is in the $100$ TeV $-$ PeV range, beyond our current capacity to directly explore. But in any rich spectrum with typical masses in this range, it is also plausible that there will be a few much lighter particles that fall within our grasp. See, e.g., the discussion in Ref.~\cite{Kilic:2009mi}.
Pseudo-Goldstone ALPs are particularly robust examples of these. From this perspective, we would expect an ALP to have all its non-renormalizable couplings to the SM characterized by these high scales, including its decay constant $f_a$ that sets its couplings to gauge bosons. Almost all such highly suppressed couplings would be phenomenologically irrelevant. But the coupling to gluons, Eq.~\eqref{aggdual}, is exceptional, as we have seen in this paper, in allowing us to sufficiently produce ALPs at the LHC given the large effective luminosity of gluons.  Such weakly coupled ALPs will be LLPs and plausibly dominated by hadronic decays. This is the same signal structure for which our proposed QCD axion search is designed.

An LLP search along the lines described in this paper will require creative designs of the triggers and analysis at the LHC main detectors.
New axion production modes can also be considered, including hadron decays and gluon splittings during parton shower. 
For lighter axions new decay modes can also be considered, ranging from exclusive modes into three pions to diphotons. 
These new production and decay considerations lead to rich phenomenology and further motivates our explorations and searches for GeV scale long-lived particles at accelerator facilities~\cite{CidVidal:2018blh,Feng:2018noy,Ariga:2018uku,Beacham:2019nyx,Aielli:2019ivi,CidVidal:2018blh,Izaguirre:2016dfi,Bauer:2017nlg,Aloni:2018vki,Dolan:2017osp,Lees:2011wb,Altmannshofer:2019yji,Marques-Tavares:2018cwm}.

{\it\bf Acknowledgments.}---
We would like to thank Matthew Daniel Citron, Jared Evans, Yuri Gershtein, Simon Knapen and Diego Redigolo for very useful comments on the draft. We would also like to thank Prateek Agrawal, Evan Berkowitz, Zohreh Davoudi and Simone Pagan Griso for helpful discussions.
This research was supported in part by the NSF grants PHY-1620074, PHY-1914480 and PHY-1914731, and by the Maryland Center for Fundamental Physics (MCFP).
AH, ZL and RS acknowledge the hospitality of the Kavli Institute for Theoretical Physics, UC Santa Barbara, during the ``Origin of the Vacuum Energy and Electroweak Scales'' workshop, and the support by the NSF grant PHY-174958.
AH and ZL would also like to thank PITT-PACC, MIAPP, and Aspen Center for Physics (supported by NSF grant PHY-1607611) for support from their programs and providing the environment for collaboration during various stages of this work. 
The codes for this study are available at \href{https://gitlab.com/ZhenLiuPhys/alplhc}{Axion@LHC}.

\onecolumngrid

\appendix
\setcounter{secnumdepth}{2}
\section{Details of the Collider Phenomenology}
\label{app:LALP}
We parametrize the coupling of the axion to the SM gauge fields as,
\begin{equation}\label{eq:alpeft}
\frac{a}{8\pi f_a}\left(c_3 \alpha_3  G\Tilde{G}+c_2\alpha_2  W\Tilde{W}+c_1\alpha_1 B\Tilde{B}\right),
\end{equation}
where $\alpha_i = \frac{g_i^2}{4\pi}$ denote the SM gauge couplings and $\alpha_1$ is related to the hypercharge gauge coupling as $\alpha_1=5/3 \alpha_Y$.
Below the scale of the electroweak symmetry breaking, the axion-photon coupling can be written as,
\begin{equation}
\frac{a}{8\pi f_a}c_\gamma \alpha_{\text{EM}} F\Tilde{F},
\end{equation}
where $c_\gamma = c_2+\frac{5}{3}c_1$ and $\alpha_{\text{EM}}=e^2/4\pi$ is the electromagnetic fine structure constant.

\subsection{Production cross section}
\label{app:production}

\begin{figure}
  \centering
  \includegraphics[scale=0.45]{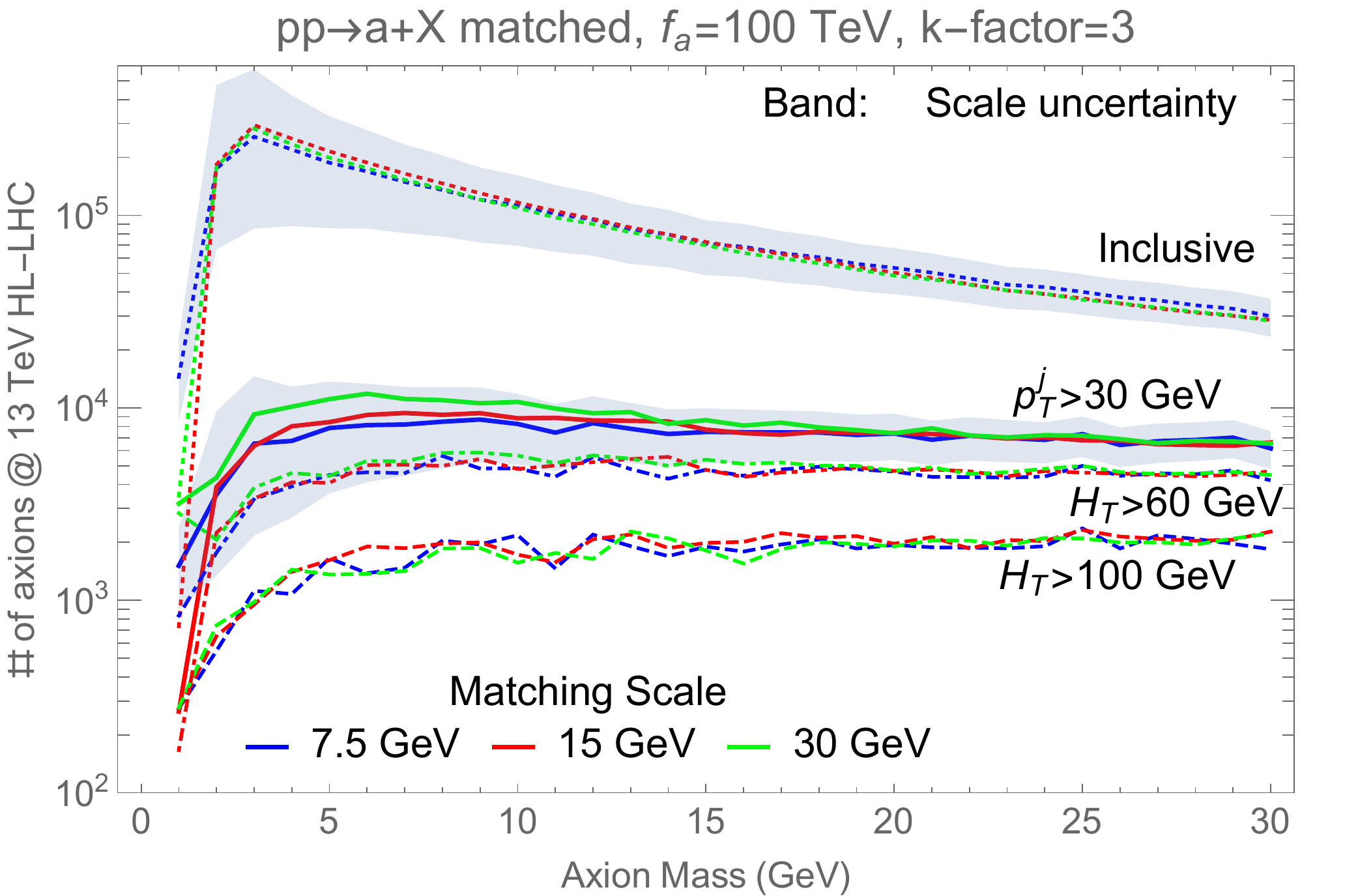}
  \caption{The number of produced axion at the 13 TeV LHC with 3 ab$^{-1}$ of integrated luminosity as a function of the axion mass in a one-jet matched cross section calculation, for a fixed axion decay constant of 100~TeV. 
  }
  \label{fig:xs}
\end{figure}

The calculation of the production cross section for the GeV scale axion, dominated by the $gg\to a$ process at leading order, is subtle. It samples the low $x$ regime and is subject to scale dependences from the running of $\alpha_3$ as well as the choices of factorization and renormalization scale. Without proper resummation, this is likely to give unphysical results in the low mass regime. Here we carry out a one-jet matched rate calculation with further detector simulation.
In Fig.~\ref{fig:xs}, we show the number of produced axion at the 13 TeV LHC with 3 ab$^{-1}$ of integrated luminosity as a function of the axion mass in a one-jet matched cross section calculation, for a fixed axion decay constant of 100~TeV. The blue, red, and green lines represent the matching scale choices of 7.5~GeV, 15~GeV, and 30~GeV, respectively. We can see that the matched calculation of the inclusive and exclusive cross section is relatively stable against the matching scale choices for axion mass above 5~GeV. The cross section un-physically decreases when the axion mass goes below 5~GeV. Similar results were also found in Ref.~\cite{CidVidal:2018blh}. The dotted line shows the inclusive cross section, and the solid, dot-dashed, and dashed lines are with a leading jet minimal $p_T$ cut of 30~GeV, a minimum $H_T$ cut of 60~GeV and 100~GeV. We can see that imposing the leading jet $p_T$ cut reduces the signal rate by one to two orders of magnitude, and the $H_T$ cut of 60~GeV and 100~GeV further reduces the rate by another factor of $\sim 1.5$ and $\sim 3.5$, respectively.
The shaded bands indicate the scale uncertainty with the matching scale choices of 7.5~GeV for the inclusive rate and the one with a minimal jet $p_T$ cut of 30~GeV, which is around $\pm 25\%$ for a broad range of the axion mass. 

The cross section calculation is carried out using {\tt Madgraph5@NLO}~\cite{Alwall:2011uj} with a modified model file based upon {\tt heft} model file, with a parton shower through {\tt Pythia8}~\cite{Sjostrand:2006za,Sjostrand:2007gs} and a further detector simulation for the response of jet clustering and $H_T$ calculation through {\tt Delphes3}~\cite{deFavereau:2013fsa}. We have also used a rough estimate of the K-factor based on Ref.~\cite{Mariotti:2017vtv}.
Details of the cross section calculation can be found in our public repository \href{https://gitlab.com/ZhenLiuPhys/alplhc}{Axion@LHC}.

\subsection{Axion decays and lifetime}
\label{app:decay}

For $m_a<3m_\pi$, the axion predominantly decays into diphotons via the coupling $a F\tilde{F}$ coming from Eq.~\eqref{eq:alpeft}, with the rate,
\begin{equation}
\Gamma_{a\rightarrow \gamma\gamma} = \frac{\alpha_{\text{EM}}^2c_\gamma^2}{256\pi^3}\frac{m_a^3}{f_a^2},
\end{equation}
while the decays $a\rightarrow \phi\gamma$ and $a\rightarrow\pi\pi$ are forbidden by C and CP invariance. 
For $m_a\gtrsim 2$ GeV, the axion predominantly decays into hadrons. The corresponding rate can be calculated from the decay rate into gluons via the coupling $a G\tilde{G}$ in Eq.~\eqref{eq:alpeft}, 
\begin{equation}
\Gamma_{a\rightarrow gg} = \frac{\alpha_3^2(m_a)c_3^2}{32 \pi^3} \frac{m_a^3}{f_a^2} K_{\text{NLO}}(m_a),
\end{equation}
where $K_{\text{NLO}}(m_a)$ is the K-factor at next-to-leading order and is given by,
\begin{equation}
K_{\text{NLO}}(m_a)=1 + \frac{\alpha_3(m_a)}{\pi} \left(\frac{97}{4} - \frac{7}{6} N_f\right).
\end{equation}
We see that the decay rate into gluons dominate over that into diphotons due to the smallness of the fine structure constant $\alpha_{\text{EM}}$ in comparison with $\alpha_3$, and the presence of color factors for the former. 
For $3m_\pi<m_a\lesssim 2$ GeV, the axion decay patterns are more complex and can be understood as from the mixings between axion and SM pseudoscalar mesons \cite{Aloni:2018vki}. New search strategies, taking into account new SM backgrounds, such as those from Kaons, need to be developed.

\begin{figure}[h]
  \centering
  \includegraphics[width=0.49\linewidth]{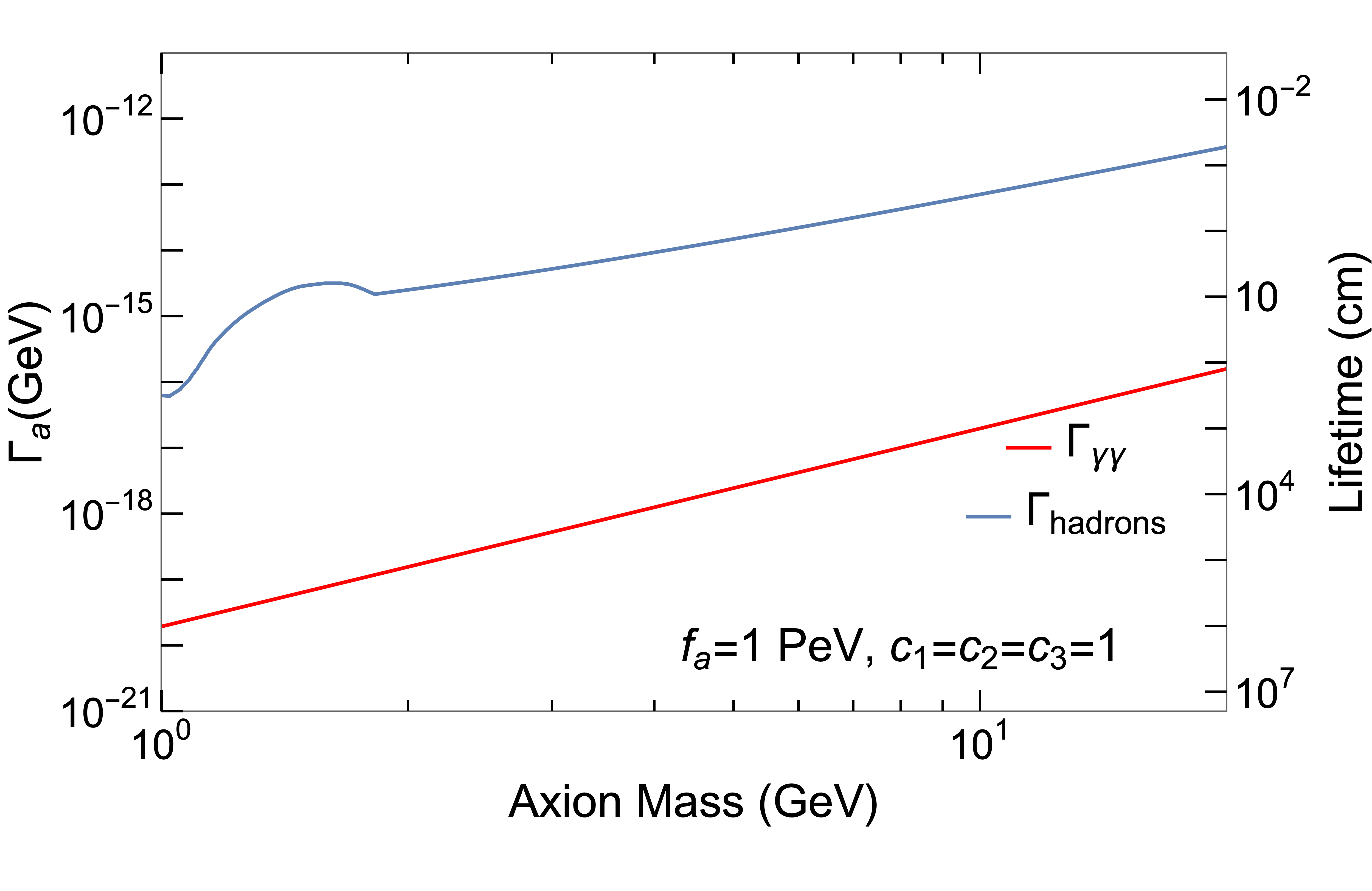}
  \caption{Axion partial decay widths into diphoton (red) and hadrons (blue) as a function of its mass. On the right side of the axes we also show the results in units of proper decay length.}
  \label{fig:width-mass}
\end{figure}

In Fig.~\ref{fig:width-mass} we show the GeV Axion partial decay widths into diphoton (red) and hadrons (blue) as a function of its mass. We chose the axion decay constant to be 1~PeV and show the corresponding inverse proper decay length on the y-axis on the right-hand side. Here, in addition to the above direct calculation, we have followed Ref.~\cite{Aloni:2018vki} in its treatment of the hadronic axion decay modes for $m_a\lesssim 2$ GeV. Such treatment generates the non-trivial behavior in the regime below 2~GeV for the $a\to g g$ calculation. The seemingly abrupt transition region near the 2 GeV regime shall be interpreted as a smooth transition between chiral perturbation treatment to matrix element calculation from the first principle. We can see that the GeV axion mainly decays into the hadrons, and the diphoton branching fraction typically is of order $10^{-3}$ to $10^{-4}$, assuming the Wilson coefficients $c_1=c_2=c_3$. This property renders the searches for axions decaying into diphotons not very sensitive to the signal, especially when it is long-lived.





\section{Current Search Coverage of GeV Axions}
\label{app:coverage}

{\tt LHCb diphoton search}-- Using 80 $\text{pb}^{-1}$ diphoton data, Ref. \cite{CidVidal:2018blh} sets a bound on $\sigma(pp\rightarrow Xa(\gamma\gamma))$ for $4.9 \gev < m_{\gamma\gamma} < 6.3 \gev$. A projection for HL phase with $300~\text{fb}^{-1}$ is also done for an extended mass range $3 \gev < m_{\gamma\gamma} < 20\gev$. We show the resulting bounds (solid)/projections (dashed) in blue for $c_1=c_2=c_3=1$ in the lower panel of Fig. \ref{fig:results}.

There are also many proposals and auxiliary experiments of the LHC that can be sensitive to sub-GeV axions. For instance, FASER, an ultra-forward detector of the LHC, would cover the axions below a GeV~\cite{Feng:2018noy,Ariga:2018uku}. More details and comparisons of the future proposals can be found in Refs.~\cite{Beacham:2019nyx,Aielli:2019ivi}. Ref. \cite{Aielli:2019ivi} has proposed that
the CODEX-b \cite{Gligorov:2017nwh} and MATHUSLA \cite{Chou:2016lxi} detector can also probe some of the long-lived regime of the parameter space.

{\tt BaBar and Belle-II}--
BaBar carried out a 
search \cite{Lees:2011wb} for $\Upsilon(2S,3S)\rightarrow\gamma a$ where $a$ decays hadronically for the mass range $2m_\pi <m_a<7 \gev$. Ref. \cite{CidVidal:2018blh} recasted this for the axion model under consideration which gives $f_a\lesssim 3.5$ GeV for $c_1=c_2=c_3=1$. A Belle-II projection was also given for the same search assuming an increase in the number of produced of $\Upsilon(3S)$ by a factor of 100 compared to BaBar and only statistical uncertainty, which gives a projection of $f_a\lesssim 10$ GeV. Assuming a future sensitivity of BR$(B\rightarrow K^{(*)}a(\gamma\gamma))= 10^{-8}$ at future B-factories, we recast the bound of Ref. \cite{Izaguirre:2016dfi}, for our benchmark point $c_1=c_2=c_3=1$. The resulting sensitivities are shown in magenta $(B\rightarrow K^* a)$ and purple ($B\rightarrow K a$) in the lower panel of Fig. \ref{fig:results}. The bounds from $B^{\pm}\rightarrow K^{\pm}a(\eta\pi^+\pi^-)$ and other hadronic decay modes of the ALP, as discussed in \cite{Aloni:2018vki}, cover a small parameter space in the lower panel of Fig. \ref{fig:results} and are not shown. We also do not show bounds from different flavor searches \cite{Dobrich:2018jyi,Gavela:2019wzg,Merlo:2019anv,Bauer:2019gfk,Cornella:2019uxs} as they do not cover a significant parameter space in Fig. \ref{fig:results}. 
The most recent proposal at Belle II searches for invisible axions and the photonic decays of visible axions bremming from photons of the decaying mesons~\cite{Dolan:2017osp}, covering the sub GeV regions. Noting that at or beyond a GeV, the hadronic decays of the axion dominates, new studies on these modes would be very promising, given the clean environment from the B-factories. For a recent study on the use of LHCb to probe signatures of hidden valleys, see Ref. \cite{Pierce:2017taw}.

{\tt Diphoton searches from the LHC and Tevatron}--
Using the measured diphoton cross section at the LHC~\cite{Aad:2012tba,Aaboud:2017vol,Chatrchyan:2014fsa}, Ref.~\cite{Mariotti:2017vtv} puts a bound on the process $\sigma(pp\rightarrow X a(\gamma\gamma))$ and projects sensitivities at HL-LHC. We show the associated limits (solid)/projections (dashed) for $c_1=c_2=c_3=1$ in brown in the lower panel of Fig.~\ref{fig:results}.

We do not show subdominant bounds coming from Z-width measurement~\cite{ALEPH:2005ab,deBlas:2016ojx}, LEP limit on BR$(Z\rightarrow \gamma a(jj))$~\cite{Adriani:1992zm}, LEP constraints on two or three photon final states \cite{Mimasu:2014nea,Jaeckel:2015jla}, heavy-ion collisions \cite{Aaboud:2017bwk,Knapen:2016moh,Knapen:2017ebd,Sirunyan:2018fhl}, ATLAS limits on BR$(Z\rightarrow\gamma a(\gamma\gamma))$~\cite{Aad:2015bua}, ATLAS search $pp\rightarrow \gamma a(\gamma\gamma)$~\cite{Aad:2015bua}, non-resonant searches for ALPs \cite{Gavela:2019cmq}, sensitivities of B-factories to $e^+e^-\rightarrow\gamma a(\gamma\gamma)$~\cite{Izaguirre:2016dfi}. For sub-GeV axions, more exclusive searches could be relevant.

\section{Analysis Details}
\label{app:kinematics}
\begin{figure}
  \centering
  \includegraphics[scale=0.5]{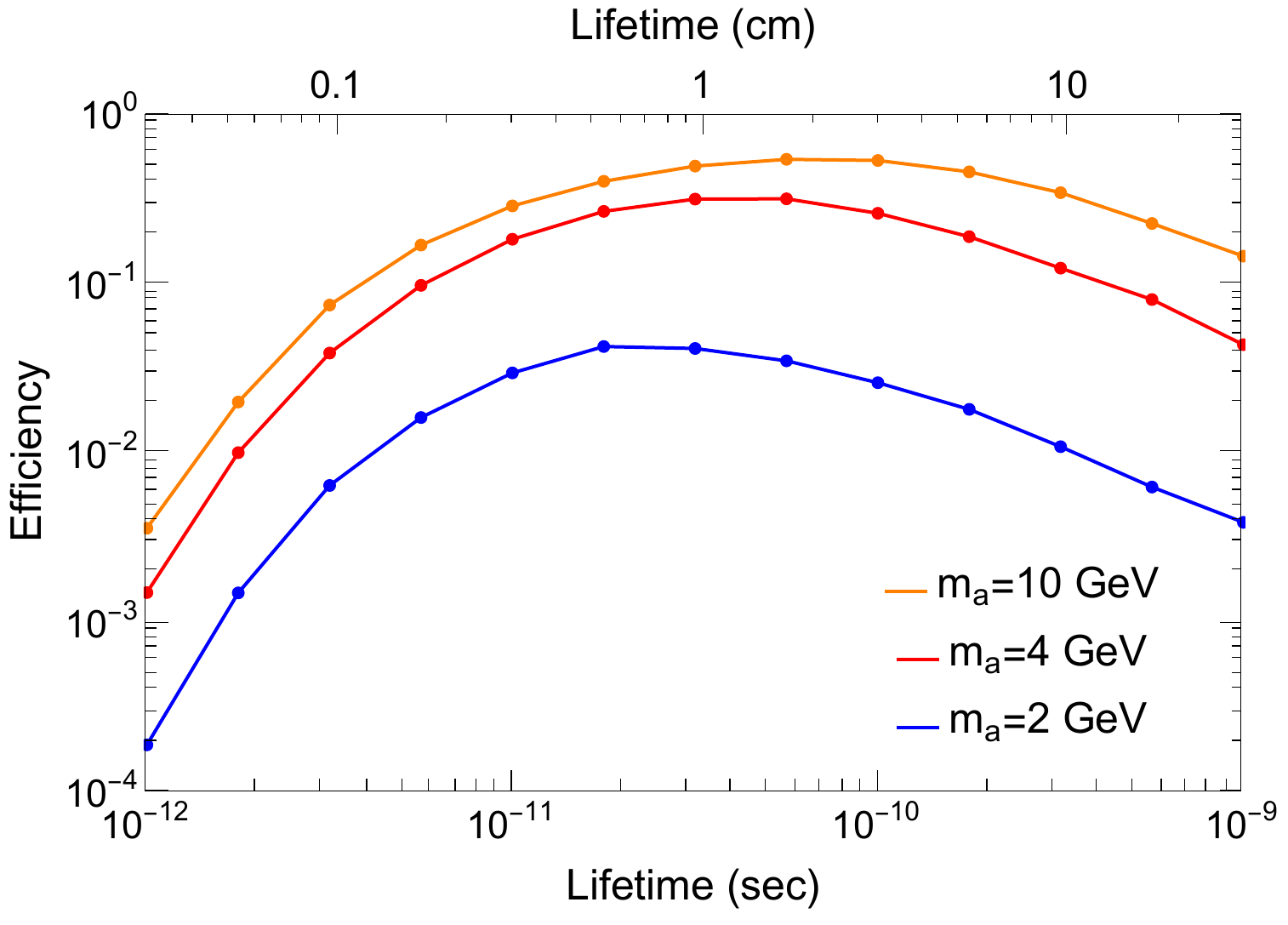}
  \caption{Selection efficiency as a function of lifetime for different masses after our selection cuts for signal samples generated with a minimum leading jet $p_T$ of $30$ GeV.
  }
  \label{fig:efflifetime}
\end{figure}
\begin{figure}
  \centering
  \includegraphics[width=1\linewidth]{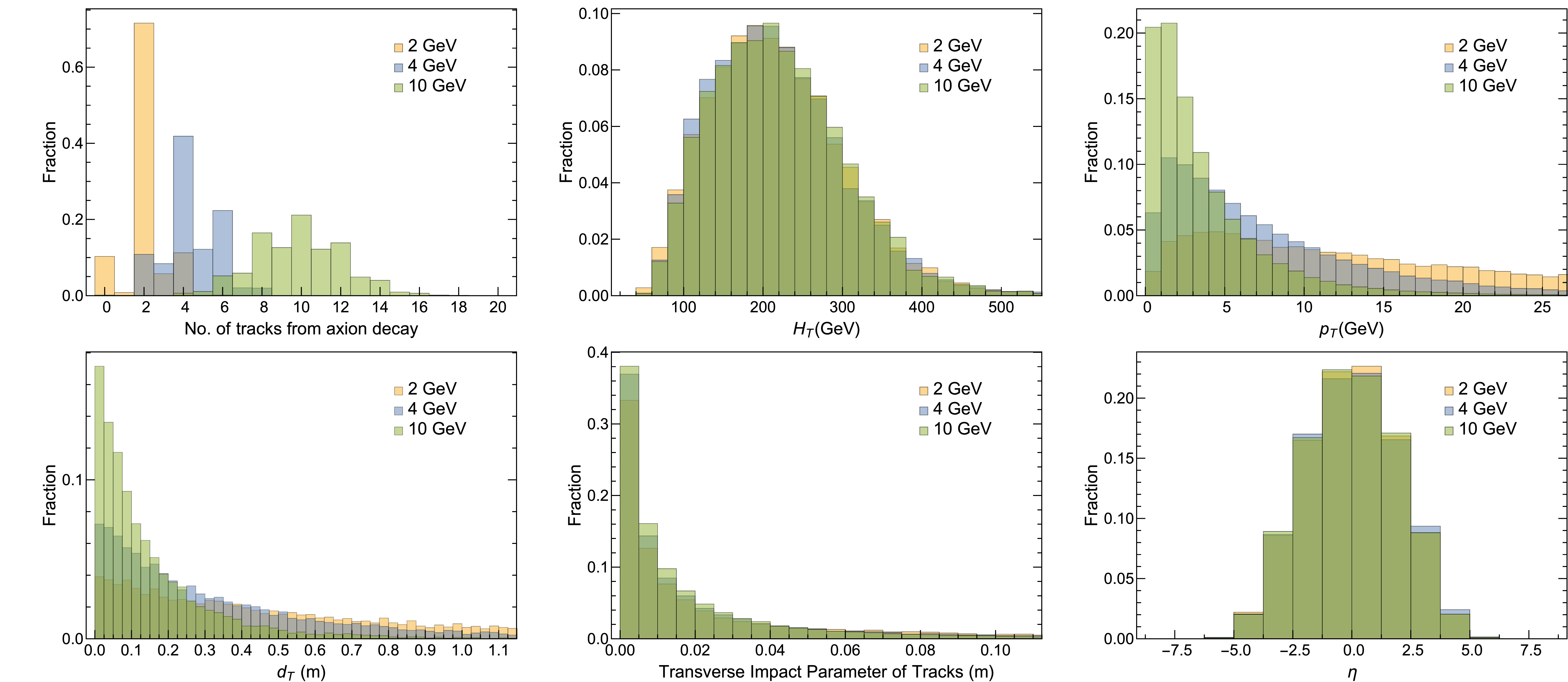}\label{fig:track}
  \caption{Relevant kinematic distributions of axion signal for three different axion masses with a common proper lifetime of $3$ cm. Clockwise from top left: (a) Number of hadronic tracks from axion decay; (b) $H_T$ distribution of the whole event; (c) $p_T$, (d) $\eta$ and (e) transverse impact parameter $d_0$ distribution of tracks from the axion decay; (f) Distribution of the decay location $d_T$ in the transverse plane.
  }
    \label{fig:nooftracks}
\end{figure}

Fig. \ref{fig:efflifetime} shows the selection efficiency as a function of axion lifetime for different axion masses.
We can see from the figure that the typical efficiency with our proposed search is $\sim 10^{-1}$ for 4~GeV and 10~GeV axion. For lighter masses, e.g., for the mass of 2~GeV shown in the blue curve, the maximum efficiency is $\sim 10^{-2}$. The peak efficiency lifetime shifts towards higher values because of the smaller boost factor.  
To get a better understanding of why lower masses have lower efficiencies, we now include the effects of various cuts in Fig.~\ref{fig:nooftracks}.


Fig.~\ref{fig:nooftracks}(a) shows the number of hadronic objects produced from the axion decay. Heavier axions produce a larger number of tracks on average. Thus it is much easier for a heavier axion to pass our requirement of three displaced tracks. 
Fig.~\ref{fig:nooftracks}(b) shows the $H_T$ distribution of jets in the entire event. We see that the requirement of having an ISR jet with $p_T>30$ GeV
at the generation level makes the resulting $H_T$ distribution largely independent of the comparatively smaller axion mass. This requirement on the ISR also leads to the feature that the $H_T$ distribution is peaked around $\sim$ 200 GeV. It is also important to note that whereas we rescaled the cross section using the reconstructed $H_T$ in Fig.~\ref{fig:xs}, in Fig.~\ref{fig:nooftracks}(b) the $H_T$ variable used is \textit{not} the reconstructed one, but rather the sum of all hadronic activities.
Fig.~\ref{fig:nooftracks}(c) shows the $p_T$ distribution of tracks coming from axion decay. Since heavier axions produce more tracks, those are also somewhat softer on average compared to tracks from lighter axions. However, we see the requirement of $p_T>2$ GeV is somewhat, but not very significant.
Figs.~\ref{fig:nooftracks}(d) and~\ref{fig:nooftracks}(e) respectively show the  $\eta$ and impact parameter distribution of tracks. The requirement of the ISR jet at the generation level again makes the results mostly independent of the axion mass. Lastly, Fig.~\ref{fig:nooftracks}(f) for the distribution of the decay location, $d_T$, in the transverse plane shows that lighter axions are more boosted on average than the heavier ones, as expected.



\section{Vertexing efficiencies and correlations}
\label{app:correlations}

\begin{figure}[tb]
  \centering
  \includegraphics[scale=0.6]{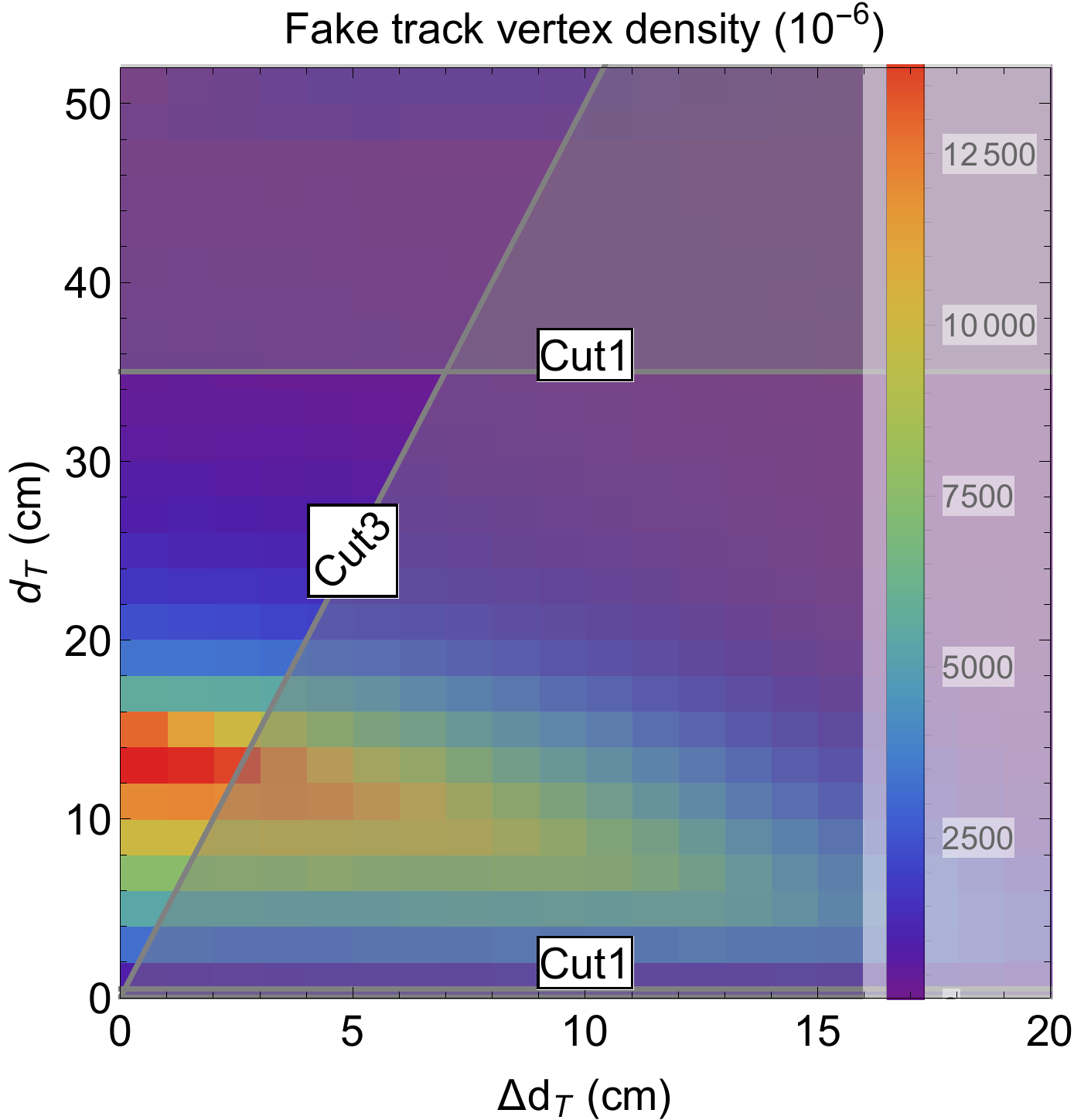}
  \caption{The probability distribution of the fake track background for the 2D-vertex reconstruction in the $\Delta d_T$-$d_T$ plane. The shaded region are excluded by our Cut1 and Cut3.
  }
  \label{fig:fakecut123}
\end{figure}

\begin{figure}
  \centering
  \includegraphics[scale=0.6]{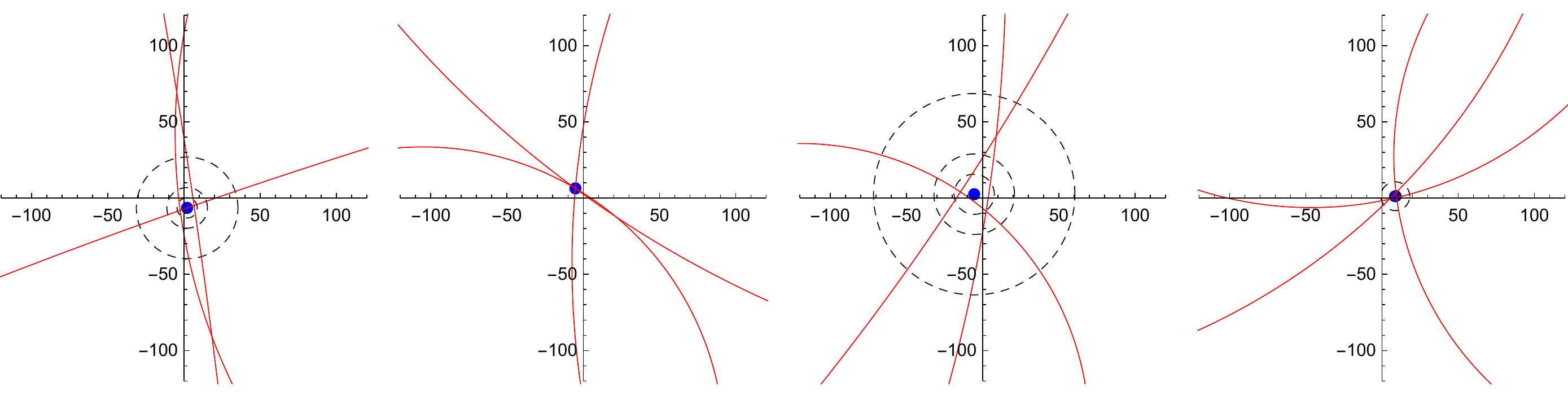}
  \caption{Typical fake track behavior in the transverse plane. The red curves represent the fake tracks. The blue dot indicates the best fit vertex location in the transverse plane. From inner to outer, the dashed circles represent the one, two, and five standard deviations of the best fit location.}
  \label{fig:fake_display}
\end{figure}

We show a few event displays of the fake-track vertexing in the transverse plane in Fig.~\ref{fig:fake_display}. The four panels represent four random instances of the one million fake-track background simulated, following our assumptions about the fake-track behavior discussed in the main text. The axes of these displays extend to the edge of the tracking volume. To obtain a more quantitative understanding of the 2D-vertex reconstruction selection, we show the distribution of the three-track vertices in Fig.~\ref{fig:fakecut123} in the $\Delta d_T$-$d_T$ plane. The shaded region is excluded by our Cut1 and Cut3. Our selection Cut2 cuts away points with $\Delta d_T>1$~cm. We can see that the fake tracks, due to the smallness of the impact parameter $d_0<15$~cm and the high $p_T$ requirement of larger than 2~GeV, do have a sizable probability to form a displaced vertex with small uncertainty in the fitted vertex location. The three-track vertex probability is 8.2\% with $\Delta d_T < 1$~cm. In fact, the 2D-cut efficiency is dominated by Cut2 as the quality of the 2D vertexing cut. Cut1 and Cut3 are chosen to make sure the selection is also effectively vetoing the SM prompt (due to smearing) and shortly long-lived backgrounds.

We further describe the correlations between various cuts considered in this study regarding the fake-track generated vertex background. Given the computational limitation, we studied one million vertices formed by the fake-tracks. A subset of the selection cuts employed in this paper would already ensure that none of the fake vertices pass. However, the HL-LHC requires a much higher suppression factor than $10^{-6}$. Hence, a study regarding the independence of many of the selection cuts and their suppression factors multiplicity is needed. 


\begin{table}[htbp]
  \centering
  \caption{Cut efficiencies on the fake track backgrounds and their correlations. The row containing $\epsilon_x$'s are the cut efficiencies for each cut (columns 1-9) and sets of cuts  in the remaining columns to the right. We abbreviate $\times 10^{-y}$ to a superscript $^{-y}$ here. Row 3-11 shows the correlations between individual cut and the cut sets, defined as in Eq.~\eqref{eq:correlation}. For positive (negative) numbers, the cuts are correlated (anti-correlated). We omit the entries with no correlations and use ``---'' to represent entries where we run out of statistics.}
    \begin{tabular}{c||c|c|c|c|c|c|c|c|c||c|c|c|c|c|c|c|c|c}
      \hline
      Cut(set) x & 1 & 2 & 3 & 4 & 5 & 6 & 7 & 8 & 9 & 1+2 & 123 & 4+5 & 6+7 & 8+9 & 4567 & 1245 & 1289 & 124567 \\
      \hline
      \hline
     $\epsilon_x$ & 9.7$^{-1}$ & 8.6$^{-2}$ & 2.3$^{-1}$ & 5.1$^{-2}$ & 6.8$^{-1}$ & 1.6$^{-2}$ & 3.1$^{-1}$ & 4.5$^{-2}$ & 1.5$^{-2}$ & 8.4$^{-2}$ & 8.2$^{-2}$ & 4.9$^{-2}$ & 3.0$^{-3}$ & 4.9$^{-4}$ & 1.6$^{-4}$ & 3.7$^{-3}$ & 1.8$^{-5}$ & 1.0$^{-5}$ \\
      \hline
      \hline
      $\rho_{1x}$ &   &   &   &  &   &   &   &   & $-$0.5 &   &   &   &   &   &   &   &   &   \\
      \hline
      $\rho_{2x}$ &   &   & +1.5 & $-$0.1 &  &   &   &   & $-$0.3 &   &   & $-$0.1 & $-$0.1 & $-$0.3 &  &   &   &   \\
      \hline
      $\rho_{3x}$ &   & +1.5 &   & $-$0.6 & $-$0.2 &  & $-$0.1 &  & +0.7 & +1.5 &   & $-$0.6 & $-$0.1 & +1.1 & $-$0.7 & +1.4 & +1.5 & +1.5 \\
      \hline
      $\rho_{4x}$ &  & $-$0.1 & $-$0.6 &   & +0.3 & +0.1 & +0.1 & +0.1 & $-$1.9 & $-$0.1 & $-$0.2 &  &  & $-$1.0 &  &   & --- &   \\
      \hline
      $\rho_{5x}$ &   &  & $-$0.2 & +0.3 &   &  &  &  & $-$0.9 &  &  &  &  & $-$0.2 &  &   & $-$0.2 &   \\
      \hline
      $\rho_{6x}$ &   &   & & +0.1 &  &   & $-$0.5 & $-$0.5 & $-$0.3 &   &   & +0.1 &  & +0.4  &  &   & ---  &   \\
      \hline
      $\rho_{7x}$ &   &   & $-$0.1 & +0.1 &  & $-$0.5 &   & +0.7 & $-$0.7 &   &   & +0.2 &  & $-$0.2 &  & +0.2 & $-$0.1  &   \\
      \hline
      $\rho_{8x}$ &   &   &  & +0.1 &  & $-$0.5 & +0.7 &   & $-$0.3 &   &   & +0.1 & +0.7 &  & +1.0 & +0.2 &  & --- \\
      \hline
      $\rho_{9x}$ & $-$0.5  & $-$0.3 & +0.7 & $-$1.9 & $-$0.9 & $-$0.3 & $-$0.7 & $-$0.3 &   & $-$0.8 & $-$0.8 & $-$2.1 & $-$0.2 &   & --- & $-$2.2 &  & --- \\
      \hline
      \end{tabular}%
  \label{tab:correlation}%
\end{table}%


In Table.~\ref{tab:correlation}, we show the individual cut efficiencies and the correlations between the different cuts employed in this study. The correlations are defined as the following, 
\beq
\rho_{AB}=\log\left(\frac {\epsilon_{AB}} {\epsilon_A \times \epsilon_B}\right),
\label{eq:correlation}
\eeq
where $\epsilon_A$ ($\epsilon_B$) is the cut efficiency of applying a set A (a set B) of cuts on the backgrounds, and $\epsilon_{AB}$ is the cut efficiency of applying the joint set of cuts $A\cap B$. 
In this study, we simulated one million fake track background and performed the vertex reconstruction on them.
Due to the limited simulation background set, it is important to check the independence of the cuts when attempting to obtain a combined cut efficiency that is more than the inverse of the number of simulated background events. 
If $\rho_{AB}=0$, it shows the two sets of cuts are completely independent. For $\rho_{AB}> 0$, the two cuts are correlated. Directly multiplying $\epsilon_A$ and $\epsilon_B$ leads to an over-estimation of the combined cut efficiencies. Similarly, for $\rho_{AB}< 0$, the two sets of cuts are anti-correlated. Directly multiplying $\epsilon_A$ and $\epsilon_B$ leads to an under-estimation of the combined cut efficiencies.
In the first row of this table, we label the abbreviated cut selections, and the second row reports the cut efficiencies for an individual cut. The columns are the corresponding correlation factor $\rho_{AB}$ for the given set of cuts. 

We can see from the Table~\ref{tab:correlation} several important correlations for use to derive the overall efficiency. First, although Cut3 provides a $2.3\times 10^{-1}$ reduction of the fake-track background, it highly correlates with Cut2 and only improves the suppression marginally, from 8.4\% to 8.2\%. Cut3 is more of a consideration to reject SM QCD background accidentally displaced due to finite detector resolution effects. We will ignore Cut3 in the overall cut chain background suppression factor calculation. The rest of the vertex reconstruction cuts, in general, have sizable anti-correlations and, at most, have a very mild correction of $+0.3$ between Cut4 and Cut5, and $+0.7$ between Cut7 to Cut8. The correlation between Cut4 and Cut5 is the consistency of the 4D vertexing information in the $z$-direction and can be eliminated when taking the cut combination of Cut4+Cut5. The correlation between Cut7 and Cut8 is offset by the anti-correlation between Cut7 and Cut9 of $-0.7$, which results in an anti-correlation between Cut7 and the combination of Cut8 and Cut9 of $-0.2$. As we can see clearly from the right-hand side of the table for the correlations between the combinations of cuts, although some entries are ``---'' due to lack of statistics, it is not hard to infer that many cut-combinations are not correlated and allows us to derive the overall cut efficiency. In this study, we take the cut combination of Cut1+Cut2+Cut8+Cut9 with background suppression of $1.8\times 10^{-5}$ and Cut4+Cut5+Cut6+Cut7 with background suppression of $1.6\times 10^{-4}$ to reach an overall estimation of the background suppression $2.9\times 10^{-9}$. Note that this choice is still conservative as Cut1+Cut2+Cut8+Cut9 mildly anti-correlates with Cut5 and Cut7, while Cut4 shows a strong anti-correlation of $-1.0$ with Cut8+Cut9, a mild anti-correlation of $-0.1$ with Cut1+Cut2, and only Cut6 has a comparatively smaller correlation with Cut8+Cut9 of $+0.4$. While we have considered a 2D-4D vertexing above, a full 4D fit will undoubtedly make our result more robust while providing higher background suppression.

\bibliography{refs_LHC_Axion}

\begin{thebibliography}{100}%
\makeatletter
\providecommand \@ifxundefined [1]{%
 \@ifx{#1\undefined}
}%
\providecommand \@ifnum [1]{%
 \ifnum #1\expandafter \@firstoftwo
 \else \expandafter \@secondoftwo
 \fi
}%
\providecommand \@ifx [1]{%
 \ifx #1\expandafter \@firstoftwo
 \else \expandafter \@secondoftwo
 \fi
}%
\providecommand \natexlab [1]{#1}%
\providecommand \enquote  [1]{``#1''}%
\providecommand \bibnamefont  [1]{#1}%
\providecommand \bibfnamefont [1]{#1}%
\providecommand \citenamefont [1]{#1}%
\providecommand \href@noop [0]{\@secondoftwo}%
\providecommand \href [0]{\begingroup \@sanitize@url \@href}%
\providecommand \@href[1]{\@@startlink{#1}\@@href}%
\providecommand \@@href[1]{\endgroup#1\@@endlink}%
\providecommand \@sanitize@url [0]{\catcode `\\12\catcode `\$12\catcode
  `\&12\catcode `\#12\catcode `\^12\catcode `\_12\catcode `\%12\relax}%
\providecommand \@@startlink[1]{}%
\providecommand \@@endlink[0]{}%
\providecommand \url  [0]{\begingroup\@sanitize@url \@url }%
\providecommand \@url [1]{\endgroup\@href {#1}{\urlprefix }}%
\providecommand \urlprefix  [0]{URL }%
\providecommand \Eprint [0]{\href }%
\providecommand \doibase [0]{http://dx.doi.org/}%
\providecommand \selectlanguage [0]{\@gobble}%
\providecommand \bibinfo  [0]{\@secondoftwo}%
\providecommand \bibfield  [0]{\@secondoftwo}%
\providecommand \translation [1]{[#1]}%
\providecommand \BibitemOpen [0]{}%
\providecommand \bibitemStop [0]{}%
\providecommand \bibitemNoStop [0]{.\EOS\space}%
\providecommand \EOS [0]{\spacefactor3000\relax}%
\providecommand \BibitemShut  [1]{\csname bibitem#1\endcsname}%
\let\auto@bib@innerbib\@empty
\bibitem [{\citenamefont {Peccei}\ and\ \citenamefont
  {Quinn}(1977{\natexlab{a}})}]{Peccei:1977hh}%
  \BibitemOpen
  \bibfield  {author} {\bibinfo {author} {\bibfnamefont {R.~D.}\ \bibnamefont
  {Peccei}}\ and\ \bibinfo {author} {\bibfnamefont {H.~R.}\ \bibnamefont
  {Quinn}},\ }\href {\doibase 10.1103/PhysRevLett.38.1440} {\bibfield
  {journal} {\bibinfo  {journal} {Phys. Rev. Lett.}\ }\textbf {\bibinfo
  {volume} {38}},\ \bibinfo {pages} {1440} (\bibinfo {year}
  {1977}{\natexlab{a}})},\ \bibinfo {note} {[,328(1977)]}\BibitemShut {NoStop}%
\bibitem [{\citenamefont {Peccei}\ and\ \citenamefont
  {Quinn}(1977{\natexlab{b}})}]{Peccei:1977ur}%
  \BibitemOpen
  \bibfield  {author} {\bibinfo {author} {\bibfnamefont {R.~D.}\ \bibnamefont
  {Peccei}}\ and\ \bibinfo {author} {\bibfnamefont {H.~R.}\ \bibnamefont
  {Quinn}},\ }\href {\doibase 10.1103/PhysRevD.16.1791} {\bibfield  {journal}
  {\bibinfo  {journal} {Phys. Rev.}\ }\textbf {\bibinfo {volume} {D16}},\
  \bibinfo {pages} {1791} (\bibinfo {year} {1977}{\natexlab{b}})}\BibitemShut
  {NoStop}%
\bibitem [{\citenamefont {Weinberg}(1978)}]{Weinberg:1977ma}%
  \BibitemOpen
  \bibfield  {author} {\bibinfo {author} {\bibfnamefont {S.}~\bibnamefont
  {Weinberg}},\ }\href {\doibase 10.1103/PhysRevLett.40.223} {\bibfield
  {journal} {\bibinfo  {journal} {Phys. Rev. Lett.}\ }\textbf {\bibinfo
  {volume} {40}},\ \bibinfo {pages} {223} (\bibinfo {year} {1978})}\BibitemShut
  {NoStop}%
\bibitem [{\citenamefont {Wilczek}(1978)}]{Wilczek:1977pj}%
  \BibitemOpen
  \bibfield  {author} {\bibinfo {author} {\bibfnamefont {F.}~\bibnamefont
  {Wilczek}},\ }\href {\doibase 10.1103/PhysRevLett.40.279} {\bibfield
  {journal} {\bibinfo  {journal} {Phys. Rev. Lett.}\ }\textbf {\bibinfo
  {volume} {40}},\ \bibinfo {pages} {279} (\bibinfo {year} {1978})}\BibitemShut
  {NoStop}%
\bibitem [{\citenamefont {Kamionkowski}\ and\ \citenamefont
  {March-Russell}(1992)}]{Kamionkowski:1992mf}%
  \BibitemOpen
  \bibfield  {author} {\bibinfo {author} {\bibfnamefont {M.}~\bibnamefont
  {Kamionkowski}}\ and\ \bibinfo {author} {\bibfnamefont {J.}~\bibnamefont
  {March-Russell}},\ }\href {\doibase 10.1016/0370-2693(92)90492-M} {\bibfield
  {journal} {\bibinfo  {journal} {Phys. Lett.}\ }\textbf {\bibinfo {volume}
  {B282}},\ \bibinfo {pages} {137} (\bibinfo {year} {1992})},\ \Eprint
  {http://arxiv.org/abs/hep-th/9202003} {arXiv:hep-th/9202003 [hep-th]}
  \BibitemShut {NoStop}%
\bibitem [{\citenamefont {Barr}\ and\ \citenamefont
  {Seckel}(1992)}]{Barr:1992qq}%
  \BibitemOpen
  \bibfield  {author} {\bibinfo {author} {\bibfnamefont {S.~M.}\ \bibnamefont
  {Barr}}\ and\ \bibinfo {author} {\bibfnamefont {D.}~\bibnamefont {Seckel}},\
  }\href {\doibase 10.1103/PhysRevD.46.539} {\bibfield  {journal} {\bibinfo
  {journal} {Phys. Rev.}\ }\textbf {\bibinfo {volume} {D46}},\ \bibinfo {pages}
  {539} (\bibinfo {year} {1992})}\BibitemShut {NoStop}%
\bibitem [{\citenamefont {Ghigna}\ \emph {et~al.}(1992)\citenamefont {Ghigna},
  \citenamefont {Lusignoli},\ and\ \citenamefont {Roncadelli}}]{GHIGNA1992278}%
  \BibitemOpen
  \bibfield  {author} {\bibinfo {author} {\bibfnamefont {S.}~\bibnamefont
  {Ghigna}}, \bibinfo {author} {\bibfnamefont {M.}~\bibnamefont {Lusignoli}}, \
  and\ \bibinfo {author} {\bibfnamefont {M.}~\bibnamefont {Roncadelli}},\
  }\href {\doibase https://doi.org/10.1016/0370-2693(92)90019-Z} {\bibfield
  {journal} {\bibinfo  {journal} {Physics Letters B}\ }\textbf {\bibinfo
  {volume} {283}},\ \bibinfo {pages} {278 } (\bibinfo {year}
  {1992})}\BibitemShut {NoStop}%
\bibitem [{\citenamefont {Holman}\ \emph {et~al.}(1992)\citenamefont {Holman},
  \citenamefont {Hsu}, \citenamefont {Kephart}, \citenamefont {Kolb},
  \citenamefont {Watkins},\ and\ \citenamefont {Widrow}}]{Holman:1992us}%
  \BibitemOpen
  \bibfield  {author} {\bibinfo {author} {\bibfnamefont {R.}~\bibnamefont
  {Holman}}, \bibinfo {author} {\bibfnamefont {S.~D.~H.}\ \bibnamefont {Hsu}},
  \bibinfo {author} {\bibfnamefont {T.~W.}\ \bibnamefont {Kephart}}, \bibinfo
  {author} {\bibfnamefont {E.~W.}\ \bibnamefont {Kolb}}, \bibinfo {author}
  {\bibfnamefont {R.}~\bibnamefont {Watkins}}, \ and\ \bibinfo {author}
  {\bibfnamefont {L.~M.}\ \bibnamefont {Widrow}},\ }\href {\doibase
  10.1016/0370-2693(92)90491-L} {\bibfield  {journal} {\bibinfo  {journal}
  {Phys. Lett.}\ }\textbf {\bibinfo {volume} {B282}},\ \bibinfo {pages} {132}
  (\bibinfo {year} {1992})},\ \Eprint {http://arxiv.org/abs/hep-ph/9203206}
  {arXiv:hep-ph/9203206 [hep-ph]} \BibitemShut {NoStop}%
\bibitem [{\citenamefont {Kim}(1985)}]{Kim:1984pt}%
  \BibitemOpen
  \bibfield  {author} {\bibinfo {author} {\bibfnamefont {J.~E.}\ \bibnamefont
  {Kim}},\ }\href {\doibase 10.1103/PhysRevD.31.1733} {\bibfield  {journal}
  {\bibinfo  {journal} {Phys. Rev.}\ }\textbf {\bibinfo {volume} {D31}},\
  \bibinfo {pages} {1733} (\bibinfo {year} {1985})}\BibitemShut {NoStop}%
\bibitem [{\citenamefont {Randall}(1992)}]{Randall:1992ut}%
  \BibitemOpen
  \bibfield  {author} {\bibinfo {author} {\bibfnamefont {L.}~\bibnamefont
  {Randall}},\ }\href {\doibase 10.1016/0370-2693(92)91928-3} {\bibfield
  {journal} {\bibinfo  {journal} {Phys. Lett.}\ }\textbf {\bibinfo {volume}
  {B284}},\ \bibinfo {pages} {77} (\bibinfo {year} {1992})}\BibitemShut
  {NoStop}%
\bibitem [{\citenamefont {Choi}(2004)}]{Choi:2003wr}%
  \BibitemOpen
  \bibfield  {author} {\bibinfo {author} {\bibfnamefont {K.-w.}\ \bibnamefont
  {Choi}},\ }\href {\doibase 10.1103/PhysRevLett.92.101602} {\bibfield
  {journal} {\bibinfo  {journal} {Phys. Rev. Lett.}\ }\textbf {\bibinfo
  {volume} {92}},\ \bibinfo {pages} {101602} (\bibinfo {year} {2004})},\
  \Eprint {http://arxiv.org/abs/hep-ph/0308024} {arXiv:hep-ph/0308024 [hep-ph]}
  \BibitemShut {NoStop}%
\bibitem [{\citenamefont {Svrcek}\ and\ \citenamefont
  {Witten}(2006)}]{Svrcek:2006yi}%
  \BibitemOpen
  \bibfield  {author} {\bibinfo {author} {\bibfnamefont {P.}~\bibnamefont
  {Svrcek}}\ and\ \bibinfo {author} {\bibfnamefont {E.}~\bibnamefont
  {Witten}},\ }\href {\doibase 10.1088/1126-6708/2006/06/051} {\bibfield
  {journal} {\bibinfo  {journal} {JHEP}\ }\textbf {\bibinfo {volume} {06}},\
  \bibinfo {pages} {051} (\bibinfo {year} {2006})},\ \Eprint
  {http://arxiv.org/abs/hep-th/0605206} {arXiv:hep-th/0605206 [hep-th]}
  \BibitemShut {NoStop}%
\bibitem [{\citenamefont {Rubakov}(1997)}]{Rubakov:1997vp}%
  \BibitemOpen
  \bibfield  {author} {\bibinfo {author} {\bibfnamefont {V.~A.}\ \bibnamefont
  {Rubakov}},\ }\href {\doibase 10.1134/1.567390} {\bibfield  {journal}
  {\bibinfo  {journal} {JETP Lett.}\ }\textbf {\bibinfo {volume} {65}},\
  \bibinfo {pages} {621} (\bibinfo {year} {1997})},\ \Eprint
  {http://arxiv.org/abs/hep-ph/9703409} {arXiv:hep-ph/9703409 [hep-ph]}
  \BibitemShut {NoStop}%
\bibitem [{\citenamefont {Berezhiani}\ \emph {et~al.}(2001)\citenamefont
  {Berezhiani}, \citenamefont {Gianfagna},\ and\ \citenamefont
  {Giannotti}}]{Berezhiani:2000gh}%
  \BibitemOpen
  \bibfield  {author} {\bibinfo {author} {\bibfnamefont {Z.}~\bibnamefont
  {Berezhiani}}, \bibinfo {author} {\bibfnamefont {L.}~\bibnamefont
  {Gianfagna}}, \ and\ \bibinfo {author} {\bibfnamefont {M.}~\bibnamefont
  {Giannotti}},\ }\href {\doibase 10.1016/S0370-2693(00)01392-7} {\bibfield
  {journal} {\bibinfo  {journal} {Phys. Lett.}\ }\textbf {\bibinfo {volume}
  {B500}},\ \bibinfo {pages} {286} (\bibinfo {year} {2001})},\ \Eprint
  {http://arxiv.org/abs/hep-ph/0009290} {arXiv:hep-ph/0009290 [hep-ph]}
  \BibitemShut {NoStop}%
\bibitem [{\citenamefont {Hook}(2015)}]{Hook:2014cda}%
  \BibitemOpen
  \bibfield  {author} {\bibinfo {author} {\bibfnamefont {A.}~\bibnamefont
  {Hook}},\ }\href {\doibase 10.1103/PhysRevLett.114.141801} {\bibfield
  {journal} {\bibinfo  {journal} {Phys. Rev. Lett.}\ }\textbf {\bibinfo
  {volume} {114}},\ \bibinfo {pages} {141801} (\bibinfo {year} {2015})},\
  \Eprint {http://arxiv.org/abs/1411.3325} {arXiv:1411.3325 [hep-ph]}
  \BibitemShut {NoStop}%
\bibitem [{\citenamefont {Fukuda}\ \emph {et~al.}(2015)\citenamefont {Fukuda},
  \citenamefont {Harigaya}, \citenamefont {Ibe},\ and\ \citenamefont
  {Yanagida}}]{Fukuda:2015ana}%
  \BibitemOpen
  \bibfield  {author} {\bibinfo {author} {\bibfnamefont {H.}~\bibnamefont
  {Fukuda}}, \bibinfo {author} {\bibfnamefont {K.}~\bibnamefont {Harigaya}},
  \bibinfo {author} {\bibfnamefont {M.}~\bibnamefont {Ibe}}, \ and\ \bibinfo
  {author} {\bibfnamefont {T.~T.}\ \bibnamefont {Yanagida}},\ }\href {\doibase
  10.1103/PhysRevD.92.015021} {\bibfield  {journal} {\bibinfo  {journal} {Phys.
  Rev.}\ }\textbf {\bibinfo {volume} {D92}},\ \bibinfo {pages} {015021}
  (\bibinfo {year} {2015})},\ \Eprint {http://arxiv.org/abs/1504.06084}
  {arXiv:1504.06084 [hep-ph]} \BibitemShut {NoStop}%
\bibitem [{\citenamefont {Dimopoulos}\ \emph {et~al.}(2016)\citenamefont
  {Dimopoulos}, \citenamefont {Hook}, \citenamefont {Huang},\ and\
  \citenamefont {Marques-Tavares}}]{Dimopoulos:2016lvn}%
  \BibitemOpen
  \bibfield  {author} {\bibinfo {author} {\bibfnamefont {S.}~\bibnamefont
  {Dimopoulos}}, \bibinfo {author} {\bibfnamefont {A.}~\bibnamefont {Hook}},
  \bibinfo {author} {\bibfnamefont {J.}~\bibnamefont {Huang}}, \ and\ \bibinfo
  {author} {\bibfnamefont {G.}~\bibnamefont {Marques-Tavares}},\ }\href
  {\doibase 10.1007/JHEP11(2016)052} {\bibfield  {journal} {\bibinfo  {journal}
  {JHEP}\ }\textbf {\bibinfo {volume} {11}},\ \bibinfo {pages} {052} (\bibinfo
  {year} {2016})},\ \Eprint {http://arxiv.org/abs/1606.03097} {arXiv:1606.03097
  [hep-ph]} \BibitemShut {NoStop}%
\bibitem [{\citenamefont {Dimopoulos}(1979)}]{Dimopoulos:1979pp}%
  \BibitemOpen
  \bibfield  {author} {\bibinfo {author} {\bibfnamefont {S.}~\bibnamefont
  {Dimopoulos}},\ }\href {\doibase 10.1016/0370-2693(79)91233-4} {\bibfield
  {journal} {\bibinfo  {journal} {Phys. Lett.}\ }\textbf {\bibinfo {volume}
  {84B}},\ \bibinfo {pages} {435} (\bibinfo {year} {1979})}\BibitemShut
  {NoStop}%
\bibitem [{\citenamefont {Holdom}\ and\ \citenamefont
  {Peskin}(1982)}]{Holdom:1982ex}%
  \BibitemOpen
  \bibfield  {author} {\bibinfo {author} {\bibfnamefont {B.}~\bibnamefont
  {Holdom}}\ and\ \bibinfo {author} {\bibfnamefont {M.~E.}\ \bibnamefont
  {Peskin}},\ }\href {\doibase 10.1016/0550-3213(82)90228-0} {\bibfield
  {journal} {\bibinfo  {journal} {Nucl. Phys.}\ }\textbf {\bibinfo {volume}
  {B208}},\ \bibinfo {pages} {397} (\bibinfo {year} {1982})}\BibitemShut
  {NoStop}%
\bibitem [{\citenamefont {Dine}\ and\ \citenamefont
  {Seiberg}(1986)}]{Dine:1986bg}%
  \BibitemOpen
  \bibfield  {author} {\bibinfo {author} {\bibfnamefont {M.}~\bibnamefont
  {Dine}}\ and\ \bibinfo {author} {\bibfnamefont {N.}~\bibnamefont {Seiberg}},\
  }\href {\doibase 10.1016/0550-3213(86)90043-X} {\bibfield  {journal}
  {\bibinfo  {journal} {Nucl. Phys.}\ }\textbf {\bibinfo {volume} {B273}},\
  \bibinfo {pages} {109} (\bibinfo {year} {1986})}\BibitemShut {NoStop}%
\bibitem [{\citenamefont {Flynn}\ and\ \citenamefont
  {Randall}(1987)}]{Flynn:1987rs}%
  \BibitemOpen
  \bibfield  {author} {\bibinfo {author} {\bibfnamefont {J.~M.}\ \bibnamefont
  {Flynn}}\ and\ \bibinfo {author} {\bibfnamefont {L.}~\bibnamefont
  {Randall}},\ }\href {\doibase 10.1016/0550-3213(87)90089-7} {\bibfield
  {journal} {\bibinfo  {journal} {Nucl. Phys.}\ }\textbf {\bibinfo {volume}
  {B293}},\ \bibinfo {pages} {731} (\bibinfo {year} {1987})}\BibitemShut
  {NoStop}%
\bibitem [{\citenamefont {Agrawal}\ and\ \citenamefont
  {Howe}(2018{\natexlab{a}})}]{Agrawal:2017ksf}%
  \BibitemOpen
  \bibfield  {author} {\bibinfo {author} {\bibfnamefont {P.}~\bibnamefont
  {Agrawal}}\ and\ \bibinfo {author} {\bibfnamefont {K.}~\bibnamefont {Howe}},\
  }\href {\doibase 10.1007/JHEP12(2018)029} {\bibfield  {journal} {\bibinfo
  {journal} {JHEP}\ }\textbf {\bibinfo {volume} {12}},\ \bibinfo {pages} {029}
  (\bibinfo {year} {2018}{\natexlab{a}})},\ \Eprint
  {http://arxiv.org/abs/1710.04213} {arXiv:1710.04213 [hep-ph]} \BibitemShut
  {NoStop}%
\bibitem [{\citenamefont {Agrawal}\ and\ \citenamefont
  {Howe}(2018{\natexlab{b}})}]{Agrawal:2017evu}%
  \BibitemOpen
  \bibfield  {author} {\bibinfo {author} {\bibfnamefont {P.}~\bibnamefont
  {Agrawal}}\ and\ \bibinfo {author} {\bibfnamefont {K.}~\bibnamefont {Howe}},\
  }\href {\doibase 10.1007/JHEP12(2018)035} {\bibfield  {journal} {\bibinfo
  {journal} {JHEP}\ }\textbf {\bibinfo {volume} {12}},\ \bibinfo {pages} {035}
  (\bibinfo {year} {2018}{\natexlab{b}})},\ \Eprint
  {http://arxiv.org/abs/1712.05803} {arXiv:1712.05803 [hep-ph]} \BibitemShut
  {NoStop}%
\bibitem [{\citenamefont {Gherghetta}\ \emph {et~al.}(2016)\citenamefont
  {Gherghetta}, \citenamefont {Nagata},\ and\ \citenamefont
  {Shifman}}]{Gherghetta:2016fhp}%
  \BibitemOpen
  \bibfield  {author} {\bibinfo {author} {\bibfnamefont {T.}~\bibnamefont
  {Gherghetta}}, \bibinfo {author} {\bibfnamefont {N.}~\bibnamefont {Nagata}},
  \ and\ \bibinfo {author} {\bibfnamefont {M.}~\bibnamefont {Shifman}},\ }\href
  {\doibase 10.1103/PhysRevD.93.115010} {\bibfield  {journal} {\bibinfo
  {journal} {Phys. Rev.}\ }\textbf {\bibinfo {volume} {D93}},\ \bibinfo {pages}
  {115010} (\bibinfo {year} {2016})},\ \Eprint
  {http://arxiv.org/abs/1604.01127} {arXiv:1604.01127 [hep-ph]} \BibitemShut
  {NoStop}%
\bibitem [{\citenamefont {Gaillard}\ \emph {et~al.}(2018)\citenamefont
  {Gaillard}, \citenamefont {Gavela}, \citenamefont {Houtz}, \citenamefont
  {Quilez},\ and\ \citenamefont {Del~Rey}}]{Gaillard:2018xgk}%
  \BibitemOpen
  \bibfield  {author} {\bibinfo {author} {\bibfnamefont {M.~K.}\ \bibnamefont
  {Gaillard}}, \bibinfo {author} {\bibfnamefont {M.~B.}\ \bibnamefont
  {Gavela}}, \bibinfo {author} {\bibfnamefont {R.}~\bibnamefont {Houtz}},
  \bibinfo {author} {\bibfnamefont {P.}~\bibnamefont {Quilez}}, \ and\ \bibinfo
  {author} {\bibfnamefont {R.}~\bibnamefont {Del~Rey}},\ }\href {\doibase
  10.1140/epjc/s10052-018-6396-6} {\bibfield  {journal} {\bibinfo  {journal}
  {Eur. Phys. J.}\ }\textbf {\bibinfo {volume} {C78}},\ \bibinfo {pages} {972}
  (\bibinfo {year} {2018})},\ \Eprint {http://arxiv.org/abs/1805.06465}
  {arXiv:1805.06465 [hep-ph]} \BibitemShut {NoStop}%
\bibitem [{\citenamefont {Graham}\ \emph {et~al.}(2015)\citenamefont {Graham},
  \citenamefont {Irastorza}, \citenamefont {Lamoreaux}, \citenamefont
  {Lindner},\ and\ \citenamefont {van Bibber}}]{Graham:2015ouw}%
  \BibitemOpen
  \bibfield  {author} {\bibinfo {author} {\bibfnamefont {P.~W.}\ \bibnamefont
  {Graham}}, \bibinfo {author} {\bibfnamefont {I.~G.}\ \bibnamefont
  {Irastorza}}, \bibinfo {author} {\bibfnamefont {S.~K.}\ \bibnamefont
  {Lamoreaux}}, \bibinfo {author} {\bibfnamefont {A.}~\bibnamefont {Lindner}},
  \ and\ \bibinfo {author} {\bibfnamefont {K.~A.}\ \bibnamefont {van Bibber}},\
  }\href {\doibase 10.1146/annurev-nucl-102014-022120} {\bibfield  {journal}
  {\bibinfo  {journal} {Ann. Rev. Nucl. Part. Sci.}\ }\textbf {\bibinfo
  {volume} {65}},\ \bibinfo {pages} {485} (\bibinfo {year} {2015})},\ \Eprint
  {http://arxiv.org/abs/1602.00039} {arXiv:1602.00039 [hep-ex]} \BibitemShut
  {NoStop}%
\bibitem [{\citenamefont {Tanabashi}\ \emph {et~al.}(2018)\citenamefont
  {Tanabashi} \emph {et~al.}}]{Tanabashi:2018oca}%
  \BibitemOpen
  \bibfield  {author} {\bibinfo {author} {\bibfnamefont {M.}~\bibnamefont
  {Tanabashi}} \emph {et~al.} (\bibinfo {collaboration} {Particle Data
  Group}),\ }\href {\doibase 10.1103/PhysRevD.98.030001} {\bibfield  {journal}
  {\bibinfo  {journal} {Phys. Rev.}\ }\textbf {\bibinfo {volume} {D98}},\
  \bibinfo {pages} {030001} (\bibinfo {year} {2018})}\BibitemShut {NoStop}%
\bibitem [{\citenamefont {Brivio}\ \emph {et~al.}(2017)\citenamefont {Brivio},
  \citenamefont {Gavela}, \citenamefont {Merlo}, \citenamefont {Mimasu},
  \citenamefont {No}, \citenamefont {del Rey},\ and\ \citenamefont
  {Sanz}}]{Brivio:2017ije}%
  \BibitemOpen
  \bibfield  {author} {\bibinfo {author} {\bibfnamefont {I.}~\bibnamefont
  {Brivio}}, \bibinfo {author} {\bibfnamefont {M.~B.}\ \bibnamefont {Gavela}},
  \bibinfo {author} {\bibfnamefont {L.}~\bibnamefont {Merlo}}, \bibinfo
  {author} {\bibfnamefont {K.}~\bibnamefont {Mimasu}}, \bibinfo {author}
  {\bibfnamefont {J.~M.}\ \bibnamefont {No}}, \bibinfo {author} {\bibfnamefont
  {R.}~\bibnamefont {del Rey}}, \ and\ \bibinfo {author} {\bibfnamefont
  {V.}~\bibnamefont {Sanz}},\ }\href {\doibase 10.1140/epjc/s10052-017-5111-3}
  {\bibfield  {journal} {\bibinfo  {journal} {Eur. Phys. J.}\ }\textbf
  {\bibinfo {volume} {C77}},\ \bibinfo {pages} {572} (\bibinfo {year}
  {2017})},\ \Eprint {http://arxiv.org/abs/1701.05379} {arXiv:1701.05379
  [hep-ph]} \BibitemShut {NoStop}%
\bibitem [{\citenamefont {Bauer}\ \emph
  {et~al.}(2017{\natexlab{a}})\citenamefont {Bauer}, \citenamefont {Neubert},\
  and\ \citenamefont {Thamm}}]{Bauer:2017ris}%
  \BibitemOpen
  \bibfield  {author} {\bibinfo {author} {\bibfnamefont {M.}~\bibnamefont
  {Bauer}}, \bibinfo {author} {\bibfnamefont {M.}~\bibnamefont {Neubert}}, \
  and\ \bibinfo {author} {\bibfnamefont {A.}~\bibnamefont {Thamm}},\ }\href
  {\doibase 10.1007/JHEP12(2017)044} {\bibfield  {journal} {\bibinfo  {journal}
  {JHEP}\ }\textbf {\bibinfo {volume} {12}},\ \bibinfo {pages} {044} (\bibinfo
  {year} {2017}{\natexlab{a}})},\ \Eprint {http://arxiv.org/abs/1708.00443}
  {arXiv:1708.00443 [hep-ph]} \BibitemShut {NoStop}%
\bibitem [{\citenamefont {Jaeckel}\ and\ \citenamefont
  {Spannowsky}(2016)}]{Jaeckel:2015jla}%
  \BibitemOpen
  \bibfield  {author} {\bibinfo {author} {\bibfnamefont {J.}~\bibnamefont
  {Jaeckel}}\ and\ \bibinfo {author} {\bibfnamefont {M.}~\bibnamefont
  {Spannowsky}},\ }\href {\doibase 10.1016/j.physletb.2015.12.037} {\bibfield
  {journal} {\bibinfo  {journal} {Phys. Lett.}\ }\textbf {\bibinfo {volume}
  {B753}},\ \bibinfo {pages} {482} (\bibinfo {year} {2016})},\ \Eprint
  {http://arxiv.org/abs/1509.00476} {arXiv:1509.00476 [hep-ph]} \BibitemShut
  {NoStop}%
\bibitem [{\citenamefont {Knapen}\ \emph {et~al.}(2017)\citenamefont {Knapen},
  \citenamefont {Lin}, \citenamefont {Lou},\ and\ \citenamefont
  {Melia}}]{Knapen:2016moh}%
  \BibitemOpen
  \bibfield  {author} {\bibinfo {author} {\bibfnamefont {S.}~\bibnamefont
  {Knapen}}, \bibinfo {author} {\bibfnamefont {T.}~\bibnamefont {Lin}},
  \bibinfo {author} {\bibfnamefont {H.~K.}\ \bibnamefont {Lou}}, \ and\
  \bibinfo {author} {\bibfnamefont {T.}~\bibnamefont {Melia}},\ }\href
  {\doibase 10.1103/PhysRevLett.118.171801} {\bibfield  {journal} {\bibinfo
  {journal} {Phys. Rev. Lett.}\ }\textbf {\bibinfo {volume} {118}},\ \bibinfo
  {pages} {171801} (\bibinfo {year} {2017})},\ \Eprint
  {http://arxiv.org/abs/1607.06083} {arXiv:1607.06083 [hep-ph]} \BibitemShut
  {NoStop}%
\bibitem [{\citenamefont {Izaguirre}\ \emph {et~al.}(2017)\citenamefont
  {Izaguirre}, \citenamefont {Lin},\ and\ \citenamefont
  {Shuve}}]{Izaguirre:2016dfi}%
  \BibitemOpen
  \bibfield  {author} {\bibinfo {author} {\bibfnamefont {E.}~\bibnamefont
  {Izaguirre}}, \bibinfo {author} {\bibfnamefont {T.}~\bibnamefont {Lin}}, \
  and\ \bibinfo {author} {\bibfnamefont {B.}~\bibnamefont {Shuve}},\ }\href
  {\doibase 10.1103/PhysRevLett.118.111802} {\bibfield  {journal} {\bibinfo
  {journal} {Phys. Rev. Lett.}\ }\textbf {\bibinfo {volume} {118}},\ \bibinfo
  {pages} {111802} (\bibinfo {year} {2017})},\ \Eprint
  {http://arxiv.org/abs/1611.09355} {arXiv:1611.09355 [hep-ph]} \BibitemShut
  {NoStop}%
\bibitem [{\citenamefont {Marciano}\ \emph {et~al.}(2016)\citenamefont
  {Marciano}, \citenamefont {Masiero}, \citenamefont {Paradisi},\ and\
  \citenamefont {Passera}}]{Marciano:2016yhf}%
  \BibitemOpen
  \bibfield  {author} {\bibinfo {author} {\bibfnamefont {W.~J.}\ \bibnamefont
  {Marciano}}, \bibinfo {author} {\bibfnamefont {A.}~\bibnamefont {Masiero}},
  \bibinfo {author} {\bibfnamefont {P.}~\bibnamefont {Paradisi}}, \ and\
  \bibinfo {author} {\bibfnamefont {M.}~\bibnamefont {Passera}},\ }\href
  {\doibase 10.1103/PhysRevD.94.115033} {\bibfield  {journal} {\bibinfo
  {journal} {Phys. Rev.}\ }\textbf {\bibinfo {volume} {D94}},\ \bibinfo {pages}
  {115033} (\bibinfo {year} {2016})},\ \Eprint
  {http://arxiv.org/abs/1607.01022} {arXiv:1607.01022 [hep-ph]} \BibitemShut
  {NoStop}%
\bibitem [{\citenamefont {Döbrich}\ \emph
  {et~al.}(2019{\natexlab{a}})\citenamefont {Döbrich}, \citenamefont {Ertas},
  \citenamefont {Kahlhoefer},\ and\ \citenamefont {Spadaro}}]{Dobrich:2018jyi}%
  \BibitemOpen
  \bibfield  {author} {\bibinfo {author} {\bibfnamefont {B.}~\bibnamefont
  {Döbrich}}, \bibinfo {author} {\bibfnamefont {F.}~\bibnamefont {Ertas}},
  \bibinfo {author} {\bibfnamefont {F.}~\bibnamefont {Kahlhoefer}}, \ and\
  \bibinfo {author} {\bibfnamefont {T.}~\bibnamefont {Spadaro}},\ }\href
  {\doibase 10.1016/j.physletb.2019.01.064} {\bibfield  {journal} {\bibinfo
  {journal} {Phys. Lett.}\ }\textbf {\bibinfo {volume} {B790}},\ \bibinfo
  {pages} {537} (\bibinfo {year} {2019}{\natexlab{a}})},\ \Eprint
  {http://arxiv.org/abs/1810.11336} {arXiv:1810.11336 [hep-ph]} \BibitemShut
  {NoStop}%
\bibitem [{\citenamefont {Craig}\ \emph {et~al.}(2018)\citenamefont {Craig},
  \citenamefont {Hook},\ and\ \citenamefont {Kasko}}]{Craig:2018kne}%
  \BibitemOpen
  \bibfield  {author} {\bibinfo {author} {\bibfnamefont {N.}~\bibnamefont
  {Craig}}, \bibinfo {author} {\bibfnamefont {A.}~\bibnamefont {Hook}}, \ and\
  \bibinfo {author} {\bibfnamefont {S.}~\bibnamefont {Kasko}},\ }\href
  {\doibase 10.1007/JHEP09(2018)028} {\bibfield  {journal} {\bibinfo  {journal}
  {JHEP}\ }\textbf {\bibinfo {volume} {09}},\ \bibinfo {pages} {028} (\bibinfo
  {year} {2018})},\ \Eprint {http://arxiv.org/abs/1805.06538} {arXiv:1805.06538
  [hep-ph]} \BibitemShut {NoStop}%
\bibitem [{\citenamefont {Gavela}\ \emph
  {et~al.}(2019{\natexlab{a}})\citenamefont {Gavela}, \citenamefont {Houtz},
  \citenamefont {Quilez}, \citenamefont {Del~Rey},\ and\ \citenamefont
  {Sumensari}}]{Gavela:2019wzg}%
  \BibitemOpen
  \bibfield  {author} {\bibinfo {author} {\bibfnamefont {M.~B.}\ \bibnamefont
  {Gavela}}, \bibinfo {author} {\bibfnamefont {R.}~\bibnamefont {Houtz}},
  \bibinfo {author} {\bibfnamefont {P.}~\bibnamefont {Quilez}}, \bibinfo
  {author} {\bibfnamefont {R.}~\bibnamefont {Del~Rey}}, \ and\ \bibinfo
  {author} {\bibfnamefont {O.}~\bibnamefont {Sumensari}},\ }\href {\doibase
  10.1140/epjc/s10052-019-6889-y} {\bibfield  {journal} {\bibinfo  {journal}
  {Eur. Phys. J.}\ }\textbf {\bibinfo {volume} {C79}},\ \bibinfo {pages} {369}
  (\bibinfo {year} {2019}{\natexlab{a}})},\ \Eprint
  {http://arxiv.org/abs/1901.02031} {arXiv:1901.02031 [hep-ph]} \BibitemShut
  {NoStop}%
\bibitem [{\citenamefont {Merlo}\ \emph {et~al.}(2019)\citenamefont {Merlo},
  \citenamefont {Pobbe}, \citenamefont {Rigolin},\ and\ \citenamefont
  {Sumensari}}]{Merlo:2019anv}%
  \BibitemOpen
  \bibfield  {author} {\bibinfo {author} {\bibfnamefont {L.}~\bibnamefont
  {Merlo}}, \bibinfo {author} {\bibfnamefont {F.}~\bibnamefont {Pobbe}},
  \bibinfo {author} {\bibfnamefont {S.}~\bibnamefont {Rigolin}}, \ and\
  \bibinfo {author} {\bibfnamefont {O.}~\bibnamefont {Sumensari}},\ }\href
  {\doibase 10.1007/JHEP06(2019)091} {\bibfield  {journal} {\bibinfo  {journal}
  {JHEP}\ }\textbf {\bibinfo {volume} {06}},\ \bibinfo {pages} {091} (\bibinfo
  {year} {2019})},\ \Eprint {http://arxiv.org/abs/1905.03259} {arXiv:1905.03259
  [hep-ph]} \BibitemShut {NoStop}%
\bibitem [{\citenamefont {Bauer}\ \emph {et~al.}(2019)\citenamefont {Bauer},
  \citenamefont {Neubert}, \citenamefont {Renner}, \citenamefont {Schnubel},\
  and\ \citenamefont {Thamm}}]{Bauer:2019gfk}%
  \BibitemOpen
  \bibfield  {author} {\bibinfo {author} {\bibfnamefont {M.}~\bibnamefont
  {Bauer}}, \bibinfo {author} {\bibfnamefont {M.}~\bibnamefont {Neubert}},
  \bibinfo {author} {\bibfnamefont {S.}~\bibnamefont {Renner}}, \bibinfo
  {author} {\bibfnamefont {M.}~\bibnamefont {Schnubel}}, \ and\ \bibinfo
  {author} {\bibfnamefont {A.}~\bibnamefont {Thamm}},\ }\href@noop {} {\
  (\bibinfo {year} {2019})},\ \Eprint {http://arxiv.org/abs/1908.00008}
  {arXiv:1908.00008 [hep-ph]} \BibitemShut {NoStop}%
\bibitem [{\citenamefont {Altmannshofer}\ \emph {et~al.}(2019)\citenamefont
  {Altmannshofer}, \citenamefont {Gori},\ and\ \citenamefont
  {Robinson}}]{Altmannshofer:2019yji}%
  \BibitemOpen
  \bibfield  {author} {\bibinfo {author} {\bibfnamefont {W.}~\bibnamefont
  {Altmannshofer}}, \bibinfo {author} {\bibfnamefont {S.}~\bibnamefont {Gori}},
  \ and\ \bibinfo {author} {\bibfnamefont {D.~J.}\ \bibnamefont {Robinson}},\
  }\href@noop {} {\  (\bibinfo {year} {2019})},\ \Eprint
  {http://arxiv.org/abs/1909.00005} {arXiv:1909.00005 [hep-ph]} \BibitemShut
  {NoStop}%
\bibitem [{\citenamefont {Jaeckel}\ \emph {et~al.}(2013)\citenamefont
  {Jaeckel}, \citenamefont {Jankowiak},\ and\ \citenamefont
  {Spannowsky}}]{Jaeckel:2012yz}%
  \BibitemOpen
  \bibfield  {author} {\bibinfo {author} {\bibfnamefont {J.}~\bibnamefont
  {Jaeckel}}, \bibinfo {author} {\bibfnamefont {M.}~\bibnamefont {Jankowiak}},
  \ and\ \bibinfo {author} {\bibfnamefont {M.}~\bibnamefont {Spannowsky}},\
  }\href {\doibase 10.1016/j.dark.2013.06.001} {\bibfield  {journal} {\bibinfo
  {journal} {Phys. Dark Univ.}\ }\textbf {\bibinfo {volume} {2}},\ \bibinfo
  {pages} {111} (\bibinfo {year} {2013})},\ \Eprint
  {http://arxiv.org/abs/1212.3620} {arXiv:1212.3620 [hep-ph]} \BibitemShut
  {NoStop}%
\bibitem [{\citenamefont {Mariotti}\ \emph {et~al.}(2018)\citenamefont
  {Mariotti}, \citenamefont {Redigolo}, \citenamefont {Sala},\ and\
  \citenamefont {Tobioka}}]{Mariotti:2017vtv}%
  \BibitemOpen
  \bibfield  {author} {\bibinfo {author} {\bibfnamefont {A.}~\bibnamefont
  {Mariotti}}, \bibinfo {author} {\bibfnamefont {D.}~\bibnamefont {Redigolo}},
  \bibinfo {author} {\bibfnamefont {F.}~\bibnamefont {Sala}}, \ and\ \bibinfo
  {author} {\bibfnamefont {K.}~\bibnamefont {Tobioka}},\ }\href {\doibase
  10.1016/j.physletb.2018.06.039} {\bibfield  {journal} {\bibinfo  {journal}
  {Phys. Lett.}\ }\textbf {\bibinfo {volume} {B783}},\ \bibinfo {pages} {13}
  (\bibinfo {year} {2018})},\ \Eprint {http://arxiv.org/abs/1710.01743}
  {arXiv:1710.01743 [hep-ph]} \BibitemShut {NoStop}%
\bibitem [{\citenamefont {Cid~Vidal}\ \emph {et~al.}(2019)\citenamefont
  {Cid~Vidal}, \citenamefont {Mariotti}, \citenamefont {Redigolo},
  \citenamefont {Sala},\ and\ \citenamefont {Tobioka}}]{CidVidal:2018blh}%
  \BibitemOpen
  \bibfield  {author} {\bibinfo {author} {\bibfnamefont {X.}~\bibnamefont
  {Cid~Vidal}}, \bibinfo {author} {\bibfnamefont {A.}~\bibnamefont {Mariotti}},
  \bibinfo {author} {\bibfnamefont {D.}~\bibnamefont {Redigolo}}, \bibinfo
  {author} {\bibfnamefont {F.}~\bibnamefont {Sala}}, \ and\ \bibinfo {author}
  {\bibfnamefont {K.}~\bibnamefont {Tobioka}},\ }\href {\doibase
  10.1007/JHEP01(2019)113} {\bibfield  {journal} {\bibinfo  {journal} {JHEP}\
  }\textbf {\bibinfo {volume} {01}},\ \bibinfo {pages} {113} (\bibinfo {year}
  {2019})},\ \Eprint {http://arxiv.org/abs/1810.09452} {arXiv:1810.09452
  [hep-ph]} \BibitemShut {NoStop}%
\bibitem [{\citenamefont {Beacham}\ \emph {et~al.}(2019)\citenamefont {Beacham}
  \emph {et~al.}}]{Beacham:2019nyx}%
  \BibitemOpen
  \bibfield  {author} {\bibinfo {author} {\bibfnamefont {J.}~\bibnamefont
  {Beacham}} \emph {et~al.},\ }\href@noop {} {\  (\bibinfo {year} {2019})},\
  \Eprint {http://arxiv.org/abs/1901.09966} {arXiv:1901.09966 [hep-ex]}
  \BibitemShut {NoStop}%
\bibitem [{\citenamefont {Aloni}\ \emph
  {et~al.}(2019{\natexlab{a}})\citenamefont {Aloni}, \citenamefont {Soreq},\
  and\ \citenamefont {Williams}}]{Aloni:2018vki}%
  \BibitemOpen
  \bibfield  {author} {\bibinfo {author} {\bibfnamefont {D.}~\bibnamefont
  {Aloni}}, \bibinfo {author} {\bibfnamefont {Y.}~\bibnamefont {Soreq}}, \ and\
  \bibinfo {author} {\bibfnamefont {M.}~\bibnamefont {Williams}},\ }\href
  {\doibase 10.1103/PhysRevLett.123.031803} {\bibfield  {journal} {\bibinfo
  {journal} {Phys. Rev. Lett.}\ }\textbf {\bibinfo {volume} {123}},\ \bibinfo
  {pages} {031803} (\bibinfo {year} {2019}{\natexlab{a}})},\ \Eprint
  {http://arxiv.org/abs/1811.03474} {arXiv:1811.03474 [hep-ph]} \BibitemShut
  {NoStop}%
\bibitem [{\citenamefont {Alonso-Álvarez}\ \emph {et~al.}(2019)\citenamefont
  {Alonso-Álvarez}, \citenamefont {Gavela},\ and\ \citenamefont
  {Quilez}}]{Alonso-Alvarez:2018irt}%
  \BibitemOpen
  \bibfield  {author} {\bibinfo {author} {\bibfnamefont {G.}~\bibnamefont
  {Alonso-Álvarez}}, \bibinfo {author} {\bibfnamefont {M.~B.}\ \bibnamefont
  {Gavela}}, \ and\ \bibinfo {author} {\bibfnamefont {P.}~\bibnamefont
  {Quilez}},\ }\href {\doibase 10.1140/epjc/s10052-019-6732-5} {\bibfield
  {journal} {\bibinfo  {journal} {Eur. Phys. J.}\ }\textbf {\bibinfo {volume}
  {C79}},\ \bibinfo {pages} {223} (\bibinfo {year} {2019})},\ \Eprint
  {http://arxiv.org/abs/1811.05466} {arXiv:1811.05466 [hep-ph]} \BibitemShut
  {NoStop}%
\bibitem [{\citenamefont {Ebadi}\ \emph {et~al.}(2019)\citenamefont {Ebadi},
  \citenamefont {Khatibi},\ and\ \citenamefont
  {Mohammadi~Najafabadi}}]{Ebadi:2019gij}%
  \BibitemOpen
  \bibfield  {author} {\bibinfo {author} {\bibfnamefont {J.}~\bibnamefont
  {Ebadi}}, \bibinfo {author} {\bibfnamefont {S.}~\bibnamefont {Khatibi}}, \
  and\ \bibinfo {author} {\bibfnamefont {M.}~\bibnamefont
  {Mohammadi~Najafabadi}},\ }\href {\doibase 10.1103/PhysRevD.100.015016}
  {\bibfield  {journal} {\bibinfo  {journal} {Phys. Rev.}\ }\textbf {\bibinfo
  {volume} {D100}},\ \bibinfo {pages} {015016} (\bibinfo {year} {2019})},\
  \Eprint {http://arxiv.org/abs/1901.03061} {arXiv:1901.03061 [hep-ph]}
  \BibitemShut {NoStop}%
\bibitem [{\citenamefont {Gavela}\ \emph
  {et~al.}(2019{\natexlab{b}})\citenamefont {Gavela}, \citenamefont {No},
  \citenamefont {Sanz},\ and\ \citenamefont {de~Trocóniz}}]{Gavela:2019cmq}%
  \BibitemOpen
  \bibfield  {author} {\bibinfo {author} {\bibfnamefont {M.~B.}\ \bibnamefont
  {Gavela}}, \bibinfo {author} {\bibfnamefont {J.~M.}\ \bibnamefont {No}},
  \bibinfo {author} {\bibfnamefont {V.}~\bibnamefont {Sanz}}, \ and\ \bibinfo
  {author} {\bibfnamefont {J.~F.}\ \bibnamefont {de~Trocóniz}},\ }\href@noop
  {} {\  (\bibinfo {year} {2019}{\natexlab{b}})},\ \Eprint
  {http://arxiv.org/abs/1905.12953} {arXiv:1905.12953 [hep-ph]} \BibitemShut
  {NoStop}%
\bibitem [{\citenamefont {Doebrich}\ \emph {et~al.}(2016)\citenamefont
  {Doebrich}, \citenamefont {Jaeckel}, \citenamefont {Kahlhoefer},
  \citenamefont {Ringwald},\ and\ \citenamefont
  {Schmidt-Hoberg}}]{Dobrich:2015jyk}%
  \BibitemOpen
  \bibfield  {author} {\bibinfo {author} {\bibfnamefont {B.}~\bibnamefont
  {Doebrich}}, \bibinfo {author} {\bibfnamefont {J.}~\bibnamefont {Jaeckel}},
  \bibinfo {author} {\bibfnamefont {F.}~\bibnamefont {Kahlhoefer}}, \bibinfo
  {author} {\bibfnamefont {A.}~\bibnamefont {Ringwald}}, \ and\ \bibinfo
  {author} {\bibfnamefont {K.}~\bibnamefont {Schmidt-Hoberg}},\ }\href
  {\doibase 10.1007/JHEP02(2016)018} {\bibfield  {journal} {\bibinfo  {journal}
  {JHEP}\ }\textbf {\bibinfo {volume} {02}},\ \bibinfo {pages} {018} (\bibinfo
  {year} {2016})},\ \Eprint {http://arxiv.org/abs/1512.03069} {arXiv:1512.03069
  [hep-ph]} \BibitemShut {NoStop}%
\bibitem [{\citenamefont {Chou}\ \emph {et~al.}(2017)\citenamefont {Chou},
  \citenamefont {Curtin},\ and\ \citenamefont {Lubatti}}]{Chou:2016lxi}%
  \BibitemOpen
  \bibfield  {author} {\bibinfo {author} {\bibfnamefont {J.~P.}\ \bibnamefont
  {Chou}}, \bibinfo {author} {\bibfnamefont {D.}~\bibnamefont {Curtin}}, \ and\
  \bibinfo {author} {\bibfnamefont {H.~J.}\ \bibnamefont {Lubatti}},\ }\href
  {\doibase 10.1016/j.physletb.2017.01.043} {\bibfield  {journal} {\bibinfo
  {journal} {Phys. Lett.}\ }\textbf {\bibinfo {volume} {B767}},\ \bibinfo
  {pages} {29} (\bibinfo {year} {2017})},\ \Eprint
  {http://arxiv.org/abs/1606.06298} {arXiv:1606.06298 [hep-ph]} \BibitemShut
  {NoStop}%
\bibitem [{\citenamefont {Dolan}\ \emph {et~al.}(2017)\citenamefont {Dolan},
  \citenamefont {Ferber}, \citenamefont {Hearty}, \citenamefont {Kahlhoefer},\
  and\ \citenamefont {Schmidt-Hoberg}}]{Dolan:2017osp}%
  \BibitemOpen
  \bibfield  {author} {\bibinfo {author} {\bibfnamefont {M.~J.}\ \bibnamefont
  {Dolan}}, \bibinfo {author} {\bibfnamefont {T.}~\bibnamefont {Ferber}},
  \bibinfo {author} {\bibfnamefont {C.}~\bibnamefont {Hearty}}, \bibinfo
  {author} {\bibfnamefont {F.}~\bibnamefont {Kahlhoefer}}, \ and\ \bibinfo
  {author} {\bibfnamefont {K.}~\bibnamefont {Schmidt-Hoberg}},\ }\href
  {\doibase 10.1007/JHEP12(2017)094} {\bibfield  {journal} {\bibinfo  {journal}
  {JHEP}\ }\textbf {\bibinfo {volume} {12}},\ \bibinfo {pages} {094} (\bibinfo
  {year} {2017})},\ \Eprint {http://arxiv.org/abs/1709.00009} {arXiv:1709.00009
  [hep-ph]} \BibitemShut {NoStop}%
\bibitem [{\citenamefont {Gligorov}\ \emph {et~al.}(2018)\citenamefont
  {Gligorov}, \citenamefont {Knapen}, \citenamefont {Papucci},\ and\
  \citenamefont {Robinson}}]{Gligorov:2017nwh}%
  \BibitemOpen
  \bibfield  {author} {\bibinfo {author} {\bibfnamefont {V.~V.}\ \bibnamefont
  {Gligorov}}, \bibinfo {author} {\bibfnamefont {S.}~\bibnamefont {Knapen}},
  \bibinfo {author} {\bibfnamefont {M.}~\bibnamefont {Papucci}}, \ and\
  \bibinfo {author} {\bibfnamefont {D.~J.}\ \bibnamefont {Robinson}},\ }\href
  {\doibase 10.1103/PhysRevD.97.015023} {\bibfield  {journal} {\bibinfo
  {journal} {Phys. Rev.}\ }\textbf {\bibinfo {volume} {D97}},\ \bibinfo {pages}
  {015023} (\bibinfo {year} {2018})},\ \Eprint
  {http://arxiv.org/abs/1708.09395} {arXiv:1708.09395 [hep-ph]} \BibitemShut
  {NoStop}%
\bibitem [{\citenamefont {Ariga}\ \emph {et~al.}(2019)\citenamefont {Ariga}
  \emph {et~al.}}]{Ariga:2018uku}%
  \BibitemOpen
  \bibfield  {author} {\bibinfo {author} {\bibfnamefont {A.}~\bibnamefont
  {Ariga}} \emph {et~al.} (\bibinfo {collaboration} {FASER}),\ }\href {\doibase
  10.1103/PhysRevD.99.095011} {\bibfield  {journal} {\bibinfo  {journal} {Phys.
  Rev.}\ }\textbf {\bibinfo {volume} {D99}},\ \bibinfo {pages} {095011}
  (\bibinfo {year} {2019})},\ \Eprint {http://arxiv.org/abs/1811.12522}
  {arXiv:1811.12522 [hep-ph]} \BibitemShut {NoStop}%
\bibitem [{\citenamefont {Berlin}\ \emph {et~al.}(2018)\citenamefont {Berlin},
  \citenamefont {Gori}, \citenamefont {Schuster},\ and\ \citenamefont
  {Toro}}]{Berlin:2018pwi}%
  \BibitemOpen
  \bibfield  {author} {\bibinfo {author} {\bibfnamefont {A.}~\bibnamefont
  {Berlin}}, \bibinfo {author} {\bibfnamefont {S.}~\bibnamefont {Gori}},
  \bibinfo {author} {\bibfnamefont {P.}~\bibnamefont {Schuster}}, \ and\
  \bibinfo {author} {\bibfnamefont {N.}~\bibnamefont {Toro}},\ }\href {\doibase
  10.1103/PhysRevD.98.035011} {\bibfield  {journal} {\bibinfo  {journal} {Phys.
  Rev.}\ }\textbf {\bibinfo {volume} {D98}},\ \bibinfo {pages} {035011}
  (\bibinfo {year} {2018})},\ \Eprint {http://arxiv.org/abs/1804.00661}
  {arXiv:1804.00661 [hep-ph]} \BibitemShut {NoStop}%
\bibitem [{\citenamefont {Feng}\ \emph {et~al.}(2018)\citenamefont {Feng},
  \citenamefont {Galon}, \citenamefont {Kling},\ and\ \citenamefont
  {Trojanowski}}]{Feng:2018noy}%
  \BibitemOpen
  \bibfield  {author} {\bibinfo {author} {\bibfnamefont {J.~L.}\ \bibnamefont
  {Feng}}, \bibinfo {author} {\bibfnamefont {I.}~\bibnamefont {Galon}},
  \bibinfo {author} {\bibfnamefont {F.}~\bibnamefont {Kling}}, \ and\ \bibinfo
  {author} {\bibfnamefont {S.}~\bibnamefont {Trojanowski}},\ }\href {\doibase
  10.1103/PhysRevD.98.055021} {\bibfield  {journal} {\bibinfo  {journal} {Phys.
  Rev.}\ }\textbf {\bibinfo {volume} {D98}},\ \bibinfo {pages} {055021}
  (\bibinfo {year} {2018})},\ \Eprint {http://arxiv.org/abs/1806.02348}
  {arXiv:1806.02348 [hep-ph]} \BibitemShut {NoStop}%
\bibitem [{\citenamefont {Curtin}\ \emph {et~al.}(2019)\citenamefont {Curtin}
  \emph {et~al.}}]{Curtin:2018mvb}%
  \BibitemOpen
  \bibfield  {author} {\bibinfo {author} {\bibfnamefont {D.}~\bibnamefont
  {Curtin}} \emph {et~al.},\ }\href {\doibase 10.1088/1361-6633/ab28d6}
  {\bibfield  {journal} {\bibinfo  {journal} {Rept. Prog. Phys.}\ }\textbf
  {\bibinfo {volume} {82}},\ \bibinfo {pages} {116201} (\bibinfo {year}
  {2019})},\ \Eprint {http://arxiv.org/abs/1806.07396} {arXiv:1806.07396
  [hep-ph]} \BibitemShut {NoStop}%
\bibitem [{\citenamefont {Aloni}\ \emph
  {et~al.}(2019{\natexlab{b}})\citenamefont {Aloni}, \citenamefont {Fanelli},
  \citenamefont {Soreq},\ and\ \citenamefont {Williams}}]{Aloni:2019ruo}%
  \BibitemOpen
  \bibfield  {author} {\bibinfo {author} {\bibfnamefont {D.}~\bibnamefont
  {Aloni}}, \bibinfo {author} {\bibfnamefont {C.}~\bibnamefont {Fanelli}},
  \bibinfo {author} {\bibfnamefont {Y.}~\bibnamefont {Soreq}}, \ and\ \bibinfo
  {author} {\bibfnamefont {M.}~\bibnamefont {Williams}},\ }\href {\doibase
  10.1103/PhysRevLett.123.071801} {\bibfield  {journal} {\bibinfo  {journal}
  {Phys. Rev. Lett.}\ }\textbf {\bibinfo {volume} {123}},\ \bibinfo {pages}
  {071801} (\bibinfo {year} {2019}{\natexlab{b}})},\ \Eprint
  {http://arxiv.org/abs/1903.03586} {arXiv:1903.03586 [hep-ph]} \BibitemShut
  {NoStop}%
\bibitem [{\citenamefont {Döbrich}\ \emph
  {et~al.}(2019{\natexlab{b}})\citenamefont {Döbrich}, \citenamefont
  {Jaeckel},\ and\ \citenamefont {Spadaro}}]{Dobrich:2019dxc}%
  \BibitemOpen
  \bibfield  {author} {\bibinfo {author} {\bibfnamefont {B.}~\bibnamefont
  {Döbrich}}, \bibinfo {author} {\bibfnamefont {J.}~\bibnamefont {Jaeckel}}, \
  and\ \bibinfo {author} {\bibfnamefont {T.}~\bibnamefont {Spadaro}},\ }\href
  {\doibase 10.1007/JHEP05(2019)213} {\bibfield  {journal} {\bibinfo  {journal}
  {JHEP}\ }\textbf {\bibinfo {volume} {05}},\ \bibinfo {pages} {213} (\bibinfo
  {year} {2019}{\natexlab{b}})},\ \Eprint {http://arxiv.org/abs/1904.02091}
  {arXiv:1904.02091 [hep-ph]} \BibitemShut {NoStop}%
\bibitem [{\citenamefont {Harland-Lang}\ \emph {et~al.}(2019)\citenamefont
  {Harland-Lang}, \citenamefont {Jaeckel},\ and\ \citenamefont
  {Spannowsky}}]{Harland-Lang:2019zur}%
  \BibitemOpen
  \bibfield  {author} {\bibinfo {author} {\bibfnamefont {L.}~\bibnamefont
  {Harland-Lang}}, \bibinfo {author} {\bibfnamefont {J.}~\bibnamefont
  {Jaeckel}}, \ and\ \bibinfo {author} {\bibfnamefont {M.}~\bibnamefont
  {Spannowsky}},\ }\href {\doibase 10.1016/j.physletb.2019.04.045} {\bibfield
  {journal} {\bibinfo  {journal} {Phys. Lett.}\ }\textbf {\bibinfo {volume}
  {B793}},\ \bibinfo {pages} {281} (\bibinfo {year} {2019})},\ \Eprint
  {http://arxiv.org/abs/1902.04878} {arXiv:1902.04878 [hep-ph]} \BibitemShut
  {NoStop}%
\bibitem [{\citenamefont {Baker}\ \emph {et~al.}(2006)\citenamefont {Baker}
  \emph {et~al.}}]{Baker:2006ts}%
  \BibitemOpen
  \bibfield  {author} {\bibinfo {author} {\bibfnamefont {C.~A.}\ \bibnamefont
  {Baker}} \emph {et~al.},\ }\href {\doibase 10.1103/PhysRevLett.97.131801}
  {\bibfield  {journal} {\bibinfo  {journal} {Phys. Rev. Lett.}\ }\textbf
  {\bibinfo {volume} {97}},\ \bibinfo {pages} {131801} (\bibinfo {year}
  {2006})},\ \Eprint {http://arxiv.org/abs/hep-ex/0602020}
  {arXiv:hep-ex/0602020 [hep-ex]} \BibitemShut {NoStop}%
\bibitem [{\citenamefont {Di~Vecchia}\ and\ \citenamefont
  {Veneziano}(1980)}]{DiVecchia:1980yfw}%
  \BibitemOpen
  \bibfield  {author} {\bibinfo {author} {\bibfnamefont {P.}~\bibnamefont
  {Di~Vecchia}}\ and\ \bibinfo {author} {\bibfnamefont {G.}~\bibnamefont
  {Veneziano}},\ }\href {\doibase 10.1016/0550-3213(80)90370-3} {\bibfield
  {journal} {\bibinfo  {journal} {Nucl. Phys.}\ }\textbf {\bibinfo {volume}
  {B171}},\ \bibinfo {pages} {253} (\bibinfo {year} {1980})}\BibitemShut
  {NoStop}%
\bibitem [{\citenamefont {Kallosh}\ \emph {et~al.}(1995)\citenamefont
  {Kallosh}, \citenamefont {Linde}, \citenamefont {Linde},\ and\ \citenamefont
  {Susskind}}]{PhysRevD.52.912}%
  \BibitemOpen
  \bibfield  {author} {\bibinfo {author} {\bibfnamefont {R.}~\bibnamefont
  {Kallosh}}, \bibinfo {author} {\bibfnamefont {A.}~\bibnamefont {Linde}},
  \bibinfo {author} {\bibfnamefont {D.}~\bibnamefont {Linde}}, \ and\ \bibinfo
  {author} {\bibfnamefont {L.}~\bibnamefont {Susskind}},\ }\href {\doibase
  10.1103/PhysRevD.52.912} {\bibfield  {journal} {\bibinfo  {journal} {Phys.
  Rev. D}\ }\textbf {\bibinfo {volume} {52}},\ \bibinfo {pages} {912} (\bibinfo
  {year} {1995})}\BibitemShut {NoStop}%
\bibitem [{\citenamefont {Banks}\ and\ \citenamefont
  {Seiberg}(2011)}]{PhysRevD.83.084019}%
  \BibitemOpen
  \bibfield  {author} {\bibinfo {author} {\bibfnamefont {T.}~\bibnamefont
  {Banks}}\ and\ \bibinfo {author} {\bibfnamefont {N.}~\bibnamefont
  {Seiberg}},\ }\href {\doibase 10.1103/PhysRevD.83.084019} {\bibfield
  {journal} {\bibinfo  {journal} {Phys. Rev. D}\ }\textbf {\bibinfo {volume}
  {83}},\ \bibinfo {pages} {084019} (\bibinfo {year} {2011})}\BibitemShut
  {NoStop}%
\bibitem [{\citenamefont {Uranga}(2007)}]{Uranga:2007zza}%
  \BibitemOpen
  \bibfield  {author} {\bibinfo {author} {\bibfnamefont {A.~M.}\ \bibnamefont
  {Uranga}},\ }\bibfield  {booktitle} {\emph {\bibinfo {booktitle} {{Strings
  and branes: The present paradigm for gauge interactions and cosmology.
  Proceedings, International Conference, Cargese School on String Theory,
  Cargese, France, May 22-June 3, 2006}}},\ }\href {\doibase
  10.1016/j.nuclphysbps.2007.06.002} {\bibfield  {journal} {\bibinfo  {journal}
  {Nucl. Phys. Proc. Suppl.}\ }\textbf {\bibinfo {volume} {171}},\ \bibinfo
  {pages} {119} (\bibinfo {year} {2007})}\BibitemShut {NoStop}%
\bibitem [{\citenamefont {Del~Debbio}\ \emph {et~al.}(2005)\citenamefont
  {Del~Debbio}, \citenamefont {Giusti},\ and\ \citenamefont
  {Pica}}]{DelDebbio:2004ns}%
  \BibitemOpen
  \bibfield  {author} {\bibinfo {author} {\bibfnamefont {L.}~\bibnamefont
  {Del~Debbio}}, \bibinfo {author} {\bibfnamefont {L.}~\bibnamefont {Giusti}},
  \ and\ \bibinfo {author} {\bibfnamefont {C.}~\bibnamefont {Pica}},\ }\href
  {\doibase 10.1103/PhysRevLett.94.032003} {\bibfield  {journal} {\bibinfo
  {journal} {Phys. Rev. Lett.}\ }\textbf {\bibinfo {volume} {94}},\ \bibinfo
  {pages} {032003} (\bibinfo {year} {2005})},\ \Eprint
  {http://arxiv.org/abs/hep-th/0407052} {arXiv:hep-th/0407052 [hep-th]}
  \BibitemShut {NoStop}%
\bibitem [{\citenamefont {Aoki}\ \emph {et~al.}(2019)\citenamefont {Aoki} \emph
  {et~al.}}]{Aoki:2019cca}%
  \BibitemOpen
  \bibfield  {author} {\bibinfo {author} {\bibfnamefont {S.}~\bibnamefont
  {Aoki}} \emph {et~al.} (\bibinfo {collaboration} {Flavour Lattice Averaging
  Group}),\ }\href@noop {} {\  (\bibinfo {year} {2019})},\ \Eprint
  {http://arxiv.org/abs/1902.08191} {arXiv:1902.08191 [hep-lat]} \BibitemShut
  {NoStop}%
\bibitem [{\citenamefont {Ellis}\ and\ \citenamefont
  {Gaillard}(1979)}]{Ellis:1978hq}%
  \BibitemOpen
  \bibfield  {author} {\bibinfo {author} {\bibfnamefont {J.~R.}\ \bibnamefont
  {Ellis}}\ and\ \bibinfo {author} {\bibfnamefont {M.~K.}\ \bibnamefont
  {Gaillard}},\ }\href {\doibase 10.1016/0550-3213(79)90297-9} {\bibfield
  {journal} {\bibinfo  {journal} {Nucl. Phys.}\ }\textbf {\bibinfo {volume}
  {B150}},\ \bibinfo {pages} {141} (\bibinfo {year} {1979})}\BibitemShut
  {NoStop}%
\bibitem [{\citenamefont {Aaboud}\ \emph
  {et~al.}(2018{\natexlab{a}})\citenamefont {Aaboud} \emph
  {et~al.}}]{Aaboud:2017nmi}%
  \BibitemOpen
  \bibfield  {author} {\bibinfo {author} {\bibfnamefont {M.}~\bibnamefont
  {Aaboud}} \emph {et~al.} (\bibinfo {collaboration} {ATLAS}),\ }\href
  {\doibase 10.1140/epjc/s10052-018-5693-4} {\bibfield  {journal} {\bibinfo
  {journal} {Eur. Phys. J.}\ }\textbf {\bibinfo {volume} {C78}},\ \bibinfo
  {pages} {250} (\bibinfo {year} {2018}{\natexlab{a}})},\ \Eprint
  {http://arxiv.org/abs/1710.07171} {arXiv:1710.07171 [hep-ex]} \BibitemShut
  {NoStop}%
\bibitem [{\citenamefont {Sirunyan}\ \emph
  {et~al.}(2018{\natexlab{a}})\citenamefont {Sirunyan} \emph
  {et~al.}}]{Sirunyan:2018xlo}%
  \BibitemOpen
  \bibfield  {author} {\bibinfo {author} {\bibfnamefont {A.~M.}\ \bibnamefont
  {Sirunyan}} \emph {et~al.} (\bibinfo {collaboration} {CMS}),\ }\href
  {\doibase 10.1007/JHEP08(2018)130} {\bibfield  {journal} {\bibinfo  {journal}
  {JHEP}\ }\textbf {\bibinfo {volume} {08}},\ \bibinfo {pages} {130} (\bibinfo
  {year} {2018}{\natexlab{a}})},\ \Eprint {http://arxiv.org/abs/1806.00843}
  {arXiv:1806.00843 [hep-ex]} \BibitemShut {NoStop}%
\bibitem [{\citenamefont {Sirunyan}\ \emph
  {et~al.}(2018{\natexlab{b}})\citenamefont {Sirunyan} \emph
  {et~al.}}]{Sirunyan:2018rlj}%
  \BibitemOpen
  \bibfield  {author} {\bibinfo {author} {\bibfnamefont {A.~M.}\ \bibnamefont
  {Sirunyan}} \emph {et~al.} (\bibinfo {collaboration} {CMS}),\ }\href
  {\doibase 10.1103/PhysRevD.98.112014} {\bibfield  {journal} {\bibinfo
  {journal} {Phys. Rev.}\ }\textbf {\bibinfo {volume} {D98}},\ \bibinfo {pages}
  {112014} (\bibinfo {year} {2018}{\natexlab{b}})},\ \Eprint
  {http://arxiv.org/abs/1808.03124} {arXiv:1808.03124 [hep-ex]} \BibitemShut
  {NoStop}%
\bibitem [{\citenamefont {Aad}\ \emph {et~al.}(2019)\citenamefont {Aad} \emph
  {et~al.}}]{Aad:2019hjw}%
  \BibitemOpen
  \bibfield  {author} {\bibinfo {author} {\bibfnamefont {G.}~\bibnamefont
  {Aad}} \emph {et~al.} (\bibinfo {collaboration} {ATLAS}),\ }\href@noop {} {\
  (\bibinfo {year} {2019})},\ \Eprint {http://arxiv.org/abs/1910.08447}
  {arXiv:1910.08447 [hep-ex]} \BibitemShut {NoStop}%
\bibitem [{\citenamefont {Gershtein}(2017)}]{Gershtein:2017tsv}%
  \BibitemOpen
  \bibfield  {author} {\bibinfo {author} {\bibfnamefont {Y.}~\bibnamefont
  {Gershtein}},\ }\href {\doibase 10.1103/PhysRevD.96.035027} {\bibfield
  {journal} {\bibinfo  {journal} {Phys. Rev.}\ }\textbf {\bibinfo {volume}
  {D96}},\ \bibinfo {pages} {035027} (\bibinfo {year} {2017})},\ \Eprint
  {http://arxiv.org/abs/1705.04321} {arXiv:1705.04321 [hep-ph]} \BibitemShut
  {NoStop}%
\bibitem [{CMS(2018)}]{CMS-PAS-FTR-18-018}%
  \BibitemOpen
  \href {https://cds.cern.ch/record/2647987} {\emph {\bibinfo {title} {{First
  Level Track Jet Trigger for Displaced Jets at High Luminosity LHC}}}},\
  \bibinfo {type} {Tech. Rep.}\ \bibinfo {number} {CMS-PAS-FTR-18-018}\
  (\bibinfo  {institution} {CERN},\ \bibinfo {address} {Geneva},\ \bibinfo
  {year} {2018})\BibitemShut {NoStop}%
\bibitem [{\citenamefont {Gershtein}\ and\ \citenamefont
  {Knapen}(2019)}]{Gershtein:2019dhy}%
  \BibitemOpen
  \bibfield  {author} {\bibinfo {author} {\bibfnamefont {Y.}~\bibnamefont
  {Gershtein}}\ and\ \bibinfo {author} {\bibfnamefont {S.}~\bibnamefont
  {Knapen}},\ }\href@noop {} {\  (\bibinfo {year} {2019})},\ \Eprint
  {http://arxiv.org/abs/1907.00007} {arXiv:1907.00007 [hep-ex]} \BibitemShut
  {NoStop}%
\bibitem [{\citenamefont {Aaboud}\ \emph
  {et~al.}(2018{\natexlab{b}})\citenamefont {Aaboud} \emph
  {et~al.}}]{Aaboud:2017iio}%
  \BibitemOpen
  \bibfield  {author} {\bibinfo {author} {\bibfnamefont {M.}~\bibnamefont
  {Aaboud}} \emph {et~al.} (\bibinfo {collaboration} {ATLAS}),\ }\href
  {\doibase 10.1103/PhysRevD.97.052012} {\bibfield  {journal} {\bibinfo
  {journal} {Phys. Rev.}\ }\textbf {\bibinfo {volume} {D97}},\ \bibinfo {pages}
  {052012} (\bibinfo {year} {2018}{\natexlab{b}})},\ \Eprint
  {http://arxiv.org/abs/1710.04901} {arXiv:1710.04901 [hep-ex]} \BibitemShut
  {NoStop}%
\bibitem [{\citenamefont {Sirunyan}\ \emph
  {et~al.}(2018{\natexlab{c}})\citenamefont {Sirunyan} \emph
  {et~al.}}]{Sirunyan:2018icq}%
  \BibitemOpen
  \bibfield  {author} {\bibinfo {author} {\bibfnamefont {A.~M.}\ \bibnamefont
  {Sirunyan}} \emph {et~al.} (\bibinfo {collaboration} {CMS}),\ }\href
  {\doibase 10.1088/1748-0221/13/10/P10034} {\bibfield  {journal} {\bibinfo
  {journal} {JINST}\ }\textbf {\bibinfo {volume} {13}},\ \bibinfo {pages}
  {P10034} (\bibinfo {year} {2018}{\natexlab{c}})},\ \Eprint
  {http://arxiv.org/abs/1807.03289} {arXiv:1807.03289 [physics.ins-det]}
  \BibitemShut {NoStop}%
\bibitem [{\citenamefont {Sirunyan}\ \emph
  {et~al.}(2019{\natexlab{a}})\citenamefont {Sirunyan} \emph
  {et~al.}}]{Sirunyan:2018vlw}%
  \BibitemOpen
  \bibfield  {author} {\bibinfo {author} {\bibfnamefont {A.~M.}\ \bibnamefont
  {Sirunyan}} \emph {et~al.} (\bibinfo {collaboration} {CMS}),\ }\href
  {\doibase 10.1103/PhysRevD.99.032011} {\bibfield  {journal} {\bibinfo
  {journal} {Phys. Rev.}\ }\textbf {\bibinfo {volume} {D99}},\ \bibinfo {pages}
  {032011} (\bibinfo {year} {2019}{\natexlab{a}})},\ \Eprint
  {http://arxiv.org/abs/1811.07991} {arXiv:1811.07991 [hep-ex]} \BibitemShut
  {NoStop}%
\bibitem [{\citenamefont {Alimena}\ \emph {et~al.}(2019)\citenamefont {Alimena}
  \emph {et~al.}}]{Alimena:2019zri}%
  \BibitemOpen
  \bibfield  {author} {\bibinfo {author} {\bibfnamefont {J.}~\bibnamefont
  {Alimena}} \emph {et~al.},\ }\href@noop {} {\  (\bibinfo {year} {2019})},\
  \Eprint {http://arxiv.org/abs/1903.04497} {arXiv:1903.04497 [hep-ex]}
  \BibitemShut {NoStop}%
\bibitem [{\citenamefont {Aghanim}\ \emph {et~al.}(2018)\citenamefont {Aghanim}
  \emph {et~al.}}]{Aghanim:2018eyx}%
  \BibitemOpen
  \bibfield  {author} {\bibinfo {author} {\bibfnamefont {N.}~\bibnamefont
  {Aghanim}} \emph {et~al.} (\bibinfo {collaboration} {Planck}),\ }\href@noop
  {} {\  (\bibinfo {year} {2018})},\ \Eprint {http://arxiv.org/abs/1807.06209}
  {arXiv:1807.06209 [astro-ph.CO]} \BibitemShut {NoStop}%
\bibitem [{\citenamefont {Kilic}\ \emph {et~al.}(2010)\citenamefont {Kilic},
  \citenamefont {Okui},\ and\ \citenamefont {Sundrum}}]{Kilic:2009mi}%
  \BibitemOpen
  \bibfield  {author} {\bibinfo {author} {\bibfnamefont {C.}~\bibnamefont
  {Kilic}}, \bibinfo {author} {\bibfnamefont {T.}~\bibnamefont {Okui}}, \ and\
  \bibinfo {author} {\bibfnamefont {R.}~\bibnamefont {Sundrum}},\ }\href
  {\doibase 10.1007/JHEP02(2010)018} {\bibfield  {journal} {\bibinfo  {journal}
  {JHEP}\ }\textbf {\bibinfo {volume} {02}},\ \bibinfo {pages} {018} (\bibinfo
  {year} {2010})},\ \Eprint {http://arxiv.org/abs/0906.0577} {arXiv:0906.0577
  [hep-ph]} \BibitemShut {NoStop}%
\bibitem [{\citenamefont {Aielli}\ \emph {et~al.}(2019)\citenamefont {Aielli}
  \emph {et~al.}}]{Aielli:2019ivi}%
  \BibitemOpen
  \bibfield  {author} {\bibinfo {author} {\bibfnamefont {G.}~\bibnamefont
  {Aielli}} \emph {et~al.},\ }\href@noop {} {\  (\bibinfo {year} {2019})},\
  \Eprint {http://arxiv.org/abs/1911.00481} {arXiv:1911.00481 [hep-ex]}
  \BibitemShut {NoStop}%
\bibitem [{\citenamefont {Bauer}\ \emph
  {et~al.}(2017{\natexlab{b}})\citenamefont {Bauer}, \citenamefont {Neubert},\
  and\ \citenamefont {Thamm}}]{Bauer:2017nlg}%
  \BibitemOpen
  \bibfield  {author} {\bibinfo {author} {\bibfnamefont {M.}~\bibnamefont
  {Bauer}}, \bibinfo {author} {\bibfnamefont {M.}~\bibnamefont {Neubert}}, \
  and\ \bibinfo {author} {\bibfnamefont {A.}~\bibnamefont {Thamm}},\ }\href
  {\doibase 10.1103/PhysRevLett.119.031802} {\bibfield  {journal} {\bibinfo
  {journal} {Phys. Rev. Lett.}\ }\textbf {\bibinfo {volume} {119}},\ \bibinfo
  {pages} {031802} (\bibinfo {year} {2017}{\natexlab{b}})},\ \Eprint
  {http://arxiv.org/abs/1704.08207} {arXiv:1704.08207 [hep-ph]} \BibitemShut
  {NoStop}%
\bibitem [{\citenamefont {Lees}\ \emph {et~al.}(2011)\citenamefont {Lees} \emph
  {et~al.}}]{Lees:2011wb}%
  \BibitemOpen
  \bibfield  {author} {\bibinfo {author} {\bibfnamefont {J.~P.}\ \bibnamefont
  {Lees}} \emph {et~al.} (\bibinfo {collaboration} {BaBar}),\ }\href {\doibase
  10.1103/PhysRevLett.107.221803} {\bibfield  {journal} {\bibinfo  {journal}
  {Phys. Rev. Lett.}\ }\textbf {\bibinfo {volume} {107}},\ \bibinfo {pages}
  {221803} (\bibinfo {year} {2011})},\ \Eprint {http://arxiv.org/abs/1108.3549}
  {arXiv:1108.3549 [hep-ex]} \BibitemShut {NoStop}%
\bibitem [{\citenamefont {Marques-Tavares}\ and\ \citenamefont
  {Teo}(2018)}]{Marques-Tavares:2018cwm}%
  \BibitemOpen
  \bibfield  {author} {\bibinfo {author} {\bibfnamefont {G.}~\bibnamefont
  {Marques-Tavares}}\ and\ \bibinfo {author} {\bibfnamefont {M.}~\bibnamefont
  {Teo}},\ }\href {\doibase 10.1007/JHEP05(2018)180} {\bibfield  {journal}
  {\bibinfo  {journal} {JHEP}\ }\textbf {\bibinfo {volume} {05}},\ \bibinfo
  {pages} {180} (\bibinfo {year} {2018})},\ \Eprint
  {http://arxiv.org/abs/1803.07575} {arXiv:1803.07575 [hep-ph]} \BibitemShut
  {NoStop}%
\bibitem [{\citenamefont {Alwall}\ \emph {et~al.}(2011)\citenamefont {Alwall},
  \citenamefont {Herquet}, \citenamefont {Maltoni}, \citenamefont {Mattelaer},\
  and\ \citenamefont {Stelzer}}]{Alwall:2011uj}%
  \BibitemOpen
  \bibfield  {author} {\bibinfo {author} {\bibfnamefont {J.}~\bibnamefont
  {Alwall}}, \bibinfo {author} {\bibfnamefont {M.}~\bibnamefont {Herquet}},
  \bibinfo {author} {\bibfnamefont {F.}~\bibnamefont {Maltoni}}, \bibinfo
  {author} {\bibfnamefont {O.}~\bibnamefont {Mattelaer}}, \ and\ \bibinfo
  {author} {\bibfnamefont {T.}~\bibnamefont {Stelzer}},\ }\href {\doibase
  10.1007/JHEP06(2011)128} {\bibfield  {journal} {\bibinfo  {journal} {JHEP}\
  }\textbf {\bibinfo {volume} {06}},\ \bibinfo {pages} {128} (\bibinfo {year}
  {2011})},\ \Eprint {http://arxiv.org/abs/1106.0522} {arXiv:1106.0522
  [hep-ph]} \BibitemShut {NoStop}%
\bibitem [{\citenamefont {Sjostrand}\ \emph {et~al.}(2006)\citenamefont
  {Sjostrand}, \citenamefont {Mrenna},\ and\ \citenamefont
  {Skands}}]{Sjostrand:2006za}%
  \BibitemOpen
  \bibfield  {author} {\bibinfo {author} {\bibfnamefont {T.}~\bibnamefont
  {Sjostrand}}, \bibinfo {author} {\bibfnamefont {S.}~\bibnamefont {Mrenna}}, \
  and\ \bibinfo {author} {\bibfnamefont {P.~Z.}\ \bibnamefont {Skands}},\
  }\href {\doibase 10.1088/1126-6708/2006/05/026} {\bibfield  {journal}
  {\bibinfo  {journal} {JHEP}\ }\textbf {\bibinfo {volume} {05}},\ \bibinfo
  {pages} {026} (\bibinfo {year} {2006})},\ \Eprint
  {http://arxiv.org/abs/hep-ph/0603175} {arXiv:hep-ph/0603175 [hep-ph]}
  \BibitemShut {NoStop}%
\bibitem [{\citenamefont {Sjostrand}\ \emph {et~al.}(2008)\citenamefont
  {Sjostrand}, \citenamefont {Mrenna},\ and\ \citenamefont
  {Skands}}]{Sjostrand:2007gs}%
  \BibitemOpen
  \bibfield  {author} {\bibinfo {author} {\bibfnamefont {T.}~\bibnamefont
  {Sjostrand}}, \bibinfo {author} {\bibfnamefont {S.}~\bibnamefont {Mrenna}}, \
  and\ \bibinfo {author} {\bibfnamefont {P.~Z.}\ \bibnamefont {Skands}},\
  }\href {\doibase 10.1016/j.cpc.2008.01.036} {\bibfield  {journal} {\bibinfo
  {journal} {Comput. Phys. Commun.}\ }\textbf {\bibinfo {volume} {178}},\
  \bibinfo {pages} {852} (\bibinfo {year} {2008})},\ \Eprint
  {http://arxiv.org/abs/0710.3820} {arXiv:0710.3820 [hep-ph]} \BibitemShut
  {NoStop}%
\bibitem [{\citenamefont {de~Favereau}\ \emph {et~al.}(2014)\citenamefont
  {de~Favereau}, \citenamefont {Delaere}, \citenamefont {Demin}, \citenamefont
  {Giammanco}, \citenamefont {Lemaître}, \citenamefont {Mertens},\ and\
  \citenamefont {Selvaggi}}]{deFavereau:2013fsa}%
  \BibitemOpen
  \bibfield  {author} {\bibinfo {author} {\bibfnamefont {J.}~\bibnamefont
  {de~Favereau}}, \bibinfo {author} {\bibfnamefont {C.}~\bibnamefont
  {Delaere}}, \bibinfo {author} {\bibfnamefont {P.}~\bibnamefont {Demin}},
  \bibinfo {author} {\bibfnamefont {A.}~\bibnamefont {Giammanco}}, \bibinfo
  {author} {\bibfnamefont {V.}~\bibnamefont {Lemaître}}, \bibinfo {author}
  {\bibfnamefont {A.}~\bibnamefont {Mertens}}, \ and\ \bibinfo {author}
  {\bibfnamefont {M.}~\bibnamefont {Selvaggi}} (\bibinfo {collaboration}
  {DELPHES 3}),\ }\href {\doibase 10.1007/JHEP02(2014)057} {\bibfield
  {journal} {\bibinfo  {journal} {JHEP}\ }\textbf {\bibinfo {volume} {02}},\
  \bibinfo {pages} {057} (\bibinfo {year} {2014})},\ \Eprint
  {http://arxiv.org/abs/1307.6346} {arXiv:1307.6346 [hep-ex]} \BibitemShut
  {NoStop}%
\bibitem [{\citenamefont {Cornella}\ \emph {et~al.}(2019)\citenamefont
  {Cornella}, \citenamefont {Paradisi},\ and\ \citenamefont
  {Sumensari}}]{Cornella:2019uxs}%
  \BibitemOpen
  \bibfield  {author} {\bibinfo {author} {\bibfnamefont {C.}~\bibnamefont
  {Cornella}}, \bibinfo {author} {\bibfnamefont {P.}~\bibnamefont {Paradisi}},
  \ and\ \bibinfo {author} {\bibfnamefont {O.}~\bibnamefont {Sumensari}},\
  }\href@noop {} {\  (\bibinfo {year} {2019})},\ \Eprint
  {http://arxiv.org/abs/1911.06279} {arXiv:1911.06279 [hep-ph]} \BibitemShut
  {NoStop}%
\bibitem [{\citenamefont {Pierce}\ \emph {et~al.}(2018)\citenamefont {Pierce},
  \citenamefont {Shakya}, \citenamefont {Tsai},\ and\ \citenamefont
  {Zhao}}]{Pierce:2017taw}%
  \BibitemOpen
  \bibfield  {author} {\bibinfo {author} {\bibfnamefont {A.}~\bibnamefont
  {Pierce}}, \bibinfo {author} {\bibfnamefont {B.}~\bibnamefont {Shakya}},
  \bibinfo {author} {\bibfnamefont {Y.}~\bibnamefont {Tsai}}, \ and\ \bibinfo
  {author} {\bibfnamefont {Y.}~\bibnamefont {Zhao}},\ }\href {\doibase
  10.1103/PhysRevD.97.095033} {\bibfield  {journal} {\bibinfo  {journal} {Phys.
  Rev.}\ }\textbf {\bibinfo {volume} {D97}},\ \bibinfo {pages} {095033}
  (\bibinfo {year} {2018})},\ \Eprint {http://arxiv.org/abs/1708.05389}
  {arXiv:1708.05389 [hep-ph]} \BibitemShut {NoStop}%
\bibitem [{\citenamefont {Aad}\ \emph {et~al.}(2013)\citenamefont {Aad} \emph
  {et~al.}}]{Aad:2012tba}%
  \BibitemOpen
  \bibfield  {author} {\bibinfo {author} {\bibfnamefont {G.}~\bibnamefont
  {Aad}} \emph {et~al.} (\bibinfo {collaboration} {ATLAS}),\ }\href {\doibase
  10.1007/JHEP01(2013)086} {\bibfield  {journal} {\bibinfo  {journal} {JHEP}\
  }\textbf {\bibinfo {volume} {01}},\ \bibinfo {pages} {086} (\bibinfo {year}
  {2013})},\ \Eprint {http://arxiv.org/abs/1211.1913} {arXiv:1211.1913
  [hep-ex]} \BibitemShut {NoStop}%
\bibitem [{\citenamefont {Aaboud}\ \emph
  {et~al.}(2017{\natexlab{a}})\citenamefont {Aaboud} \emph
  {et~al.}}]{Aaboud:2017vol}%
  \BibitemOpen
  \bibfield  {author} {\bibinfo {author} {\bibfnamefont {M.}~\bibnamefont
  {Aaboud}} \emph {et~al.} (\bibinfo {collaboration} {ATLAS}),\ }\href
  {\doibase 10.1103/PhysRevD.95.112005} {\bibfield  {journal} {\bibinfo
  {journal} {Phys. Rev.}\ }\textbf {\bibinfo {volume} {D95}},\ \bibinfo {pages}
  {112005} (\bibinfo {year} {2017}{\natexlab{a}})},\ \Eprint
  {http://arxiv.org/abs/1704.03839} {arXiv:1704.03839 [hep-ex]} \BibitemShut
  {NoStop}%
\bibitem [{\citenamefont {Chatrchyan}\ \emph {et~al.}(2014)\citenamefont
  {Chatrchyan} \emph {et~al.}}]{Chatrchyan:2014fsa}%
  \BibitemOpen
  \bibfield  {author} {\bibinfo {author} {\bibfnamefont {S.}~\bibnamefont
  {Chatrchyan}} \emph {et~al.} (\bibinfo {collaboration} {CMS}),\ }\href
  {\doibase 10.1140/epjc/s10052-014-3129-3} {\bibfield  {journal} {\bibinfo
  {journal} {Eur. Phys. J.}\ }\textbf {\bibinfo {volume} {C74}},\ \bibinfo
  {pages} {3129} (\bibinfo {year} {2014})},\ \Eprint
  {http://arxiv.org/abs/1405.7225} {arXiv:1405.7225 [hep-ex]} \BibitemShut
  {NoStop}%
\bibitem [{\citenamefont {Schael}\ \emph {et~al.}(2006)\citenamefont {Schael}
  \emph {et~al.}}]{ALEPH:2005ab}%
  \BibitemOpen
  \bibfield  {author} {\bibinfo {author} {\bibfnamefont {S.}~\bibnamefont
  {Schael}} \emph {et~al.} (\bibinfo {collaboration} {ALEPH, DELPHI, L3, OPAL,
  SLD, LEP Electroweak Working Group, SLD Electroweak Group, SLD Heavy Flavour
  Group}),\ }\href {\doibase 10.1016/j.physrep.2005.12.006} {\bibfield
  {journal} {\bibinfo  {journal} {Phys. Rept.}\ }\textbf {\bibinfo {volume}
  {427}},\ \bibinfo {pages} {257} (\bibinfo {year} {2006})},\ \Eprint
  {http://arxiv.org/abs/hep-ex/0509008} {arXiv:hep-ex/0509008 [hep-ex]}
  \BibitemShut {NoStop}%
\bibitem [{\citenamefont {de~Blas}\ \emph {et~al.}(2016)\citenamefont
  {de~Blas}, \citenamefont {Ciuchini}, \citenamefont {Franco}, \citenamefont
  {Mishima}, \citenamefont {Pierini}, \citenamefont {Reina},\ and\
  \citenamefont {Silvestrini}}]{deBlas:2016ojx}%
  \BibitemOpen
  \bibfield  {author} {\bibinfo {author} {\bibfnamefont {J.}~\bibnamefont
  {de~Blas}}, \bibinfo {author} {\bibfnamefont {M.}~\bibnamefont {Ciuchini}},
  \bibinfo {author} {\bibfnamefont {E.}~\bibnamefont {Franco}}, \bibinfo
  {author} {\bibfnamefont {S.}~\bibnamefont {Mishima}}, \bibinfo {author}
  {\bibfnamefont {M.}~\bibnamefont {Pierini}}, \bibinfo {author} {\bibfnamefont
  {L.}~\bibnamefont {Reina}}, \ and\ \bibinfo {author} {\bibfnamefont
  {L.}~\bibnamefont {Silvestrini}},\ }\href {\doibase 10.1007/JHEP12(2016)135}
  {\bibfield  {journal} {\bibinfo  {journal} {JHEP}\ }\textbf {\bibinfo
  {volume} {12}},\ \bibinfo {pages} {135} (\bibinfo {year} {2016})},\ \Eprint
  {http://arxiv.org/abs/1608.01509} {arXiv:1608.01509 [hep-ph]} \BibitemShut
  {NoStop}%
\bibitem [{\citenamefont {Adriani}\ \emph {et~al.}(1992)\citenamefont {Adriani}
  \emph {et~al.}}]{Adriani:1992zm}%
  \BibitemOpen
  \bibfield  {author} {\bibinfo {author} {\bibfnamefont {O.}~\bibnamefont
  {Adriani}} \emph {et~al.} (\bibinfo {collaboration} {L3}),\ }\href {\doibase
  10.1016/0370-2693(92)91205-N} {\bibfield  {journal} {\bibinfo  {journal}
  {Phys. Lett.}\ }\textbf {\bibinfo {volume} {B292}},\ \bibinfo {pages} {472}
  (\bibinfo {year} {1992})}\BibitemShut {NoStop}%
\bibitem [{\citenamefont {Mimasu}\ and\ \citenamefont
  {Sanz}(2015)}]{Mimasu:2014nea}%
  \BibitemOpen
  \bibfield  {author} {\bibinfo {author} {\bibfnamefont {K.}~\bibnamefont
  {Mimasu}}\ and\ \bibinfo {author} {\bibfnamefont {V.}~\bibnamefont {Sanz}},\
  }\href {\doibase 10.1007/JHEP06(2015)173} {\bibfield  {journal} {\bibinfo
  {journal} {JHEP}\ }\textbf {\bibinfo {volume} {06}},\ \bibinfo {pages} {173}
  (\bibinfo {year} {2015})},\ \Eprint {http://arxiv.org/abs/1409.4792}
  {arXiv:1409.4792 [hep-ph]} \BibitemShut {NoStop}%
\bibitem [{\citenamefont {Aaboud}\ \emph
  {et~al.}(2017{\natexlab{b}})\citenamefont {Aaboud} \emph
  {et~al.}}]{Aaboud:2017bwk}%
  \BibitemOpen
  \bibfield  {author} {\bibinfo {author} {\bibfnamefont {M.}~\bibnamefont
  {Aaboud}} \emph {et~al.} (\bibinfo {collaboration} {ATLAS}),\ }\href
  {\doibase 10.1038/nphys4208} {\bibfield  {journal} {\bibinfo  {journal}
  {Nature Phys.}\ }\textbf {\bibinfo {volume} {13}},\ \bibinfo {pages} {852}
  (\bibinfo {year} {2017}{\natexlab{b}})},\ \Eprint
  {http://arxiv.org/abs/1702.01625} {arXiv:1702.01625 [hep-ex]} \BibitemShut
  {NoStop}%
\bibitem [{\citenamefont {Knapen}\ \emph {et~al.}(2018)\citenamefont {Knapen},
  \citenamefont {Lin}, \citenamefont {Lou},\ and\ \citenamefont
  {Melia}}]{Knapen:2017ebd}%
  \BibitemOpen
  \bibfield  {author} {\bibinfo {author} {\bibfnamefont {S.}~\bibnamefont
  {Knapen}}, \bibinfo {author} {\bibfnamefont {T.}~\bibnamefont {Lin}},
  \bibinfo {author} {\bibfnamefont {H.~K.}\ \bibnamefont {Lou}}, \ and\
  \bibinfo {author} {\bibfnamefont {T.}~\bibnamefont {Melia}},\ }\bibfield
  {booktitle} {\emph {\bibinfo {booktitle} {{Proceedings of the PHOTON-2017
  Conference: CERN, Geneva, Switzerland, May 22-26, 2017}}},\ }\href {\doibase
  10.23727/CERN-Proceedings-2018-001.65} {\bibfield  {journal} {\bibinfo
  {journal} {CERN Proc.}\ }\textbf {\bibinfo {volume} {1}},\ \bibinfo {pages}
  {65} (\bibinfo {year} {2018})},\ \Eprint {http://arxiv.org/abs/1709.07110}
  {arXiv:1709.07110 [hep-ph]} \BibitemShut {NoStop}%
\bibitem [{\citenamefont {Sirunyan}\ \emph
  {et~al.}(2019{\natexlab{b}})\citenamefont {Sirunyan} \emph
  {et~al.}}]{Sirunyan:2018fhl}%
  \BibitemOpen
  \bibfield  {author} {\bibinfo {author} {\bibfnamefont {A.~M.}\ \bibnamefont
  {Sirunyan}} \emph {et~al.} (\bibinfo {collaboration} {CMS}),\ }\href
  {\doibase 10.1016/j.physletb.2019.134826} {\bibfield  {journal} {\bibinfo
  {journal} {Phys. Lett.}\ }\textbf {\bibinfo {volume} {B797}},\ \bibinfo
  {pages} {134826} (\bibinfo {year} {2019}{\natexlab{b}})},\ \Eprint
  {http://arxiv.org/abs/1810.04602} {arXiv:1810.04602 [hep-ex]} \BibitemShut
  {NoStop}%
\bibitem [{\citenamefont {Aad}\ \emph {et~al.}(2016)\citenamefont {Aad} \emph
  {et~al.}}]{Aad:2015bua}%
  \BibitemOpen
  \bibfield  {author} {\bibinfo {author} {\bibfnamefont {G.}~\bibnamefont
  {Aad}} \emph {et~al.} (\bibinfo {collaboration} {ATLAS}),\ }\href {\doibase
  10.1140/epjc/s10052-016-4034-8} {\bibfield  {journal} {\bibinfo  {journal}
  {Eur. Phys. J.}\ }\textbf {\bibinfo {volume} {C76}},\ \bibinfo {pages} {210}
  (\bibinfo {year} {2016})},\ \Eprint {http://arxiv.org/abs/1509.05051}
  {arXiv:1509.05051 [hep-ex]} \BibitemShut {NoStop}%
\end{thebibliography}%

\end{document}